\documentclass[11pt,a4paper]{article}
\usepackage{jheppub}

\usepackage{subfigure}
\usepackage{cancel}
\usepackage{here}
\usepackage{comment,braket}
\usepackage{amsmath}
\usepackage{graphics}
\usepackage{amsfonts}
\usepackage{amssymb}
\usepackage{latexsym}
\usepackage{graphicx}

\preprint{CTPU-PTC-22-18}

\title{Lepton Flavour Violation Tests of Type II Seesaw Leptogenesis}
\author[a]{N. D. Barrie}
\author[,b,c]{and S. T. Petcov\footnote{Also at: Institute of Nuclear Research and Nuclear Energy, Bulgarian Academy of Sciences, 1784 Sofia,
Bulgaria.}}
\affiliation[a]{Center for Theoretical Physics of the Universe, Institute for Basic Science (IBS),\\
		Daejeon, 34126, Korea.}
\affiliation[b]{SISSA/INFN, Via Bonomea 265, 34136 Trieste, Italy.}
\affiliation[c]{Kavli IPMU (WPI), UTIAS,
The University of Tokyo, Kashiwa, Chiba 277-8583, Japan.}

\emailAdd{nlbarrie@ibs.re.kr}
\emailAdd{petcov@sissa.it}
\abstract{
Upcoming Lepton Flavour Violation experiments searching for $\mu\rightarrow 3e$ and $\mu$ to $ e $ conversion in nuclei processes will provide new opportunities to test the fundamental properties of the neutrino sector, and possibly the origin of matter. In recent work, it was shown that the Type II Seesaw mechanism alone can simultaneously explain the neutrino masses, Leptogenesis, and inflation. A key prediction of this model was the possibility of signals being produced in Lepton Flavour Violation decays. Searches at future experiments such as Mu3e and COMET will be integral to determining the properties of the associated triplet Higgs, and will complement other terrestrial experimental searches and cosmological measurements. In this work, we survey the detection prospects for the ingredients of the Type II Seesaw Leptogenesis scenario, and discuss the corresponding dependencies on the neutrino oscillation parameters and $\mathcal{CP}$ phases.}

\arxivnumber{2210.XXXXX}

\begin{document}
\maketitle

\section{Introduction}
\label{intro}

The existence of non-zero neutrino masses is a fundamental mystery of modern physics. Many extensions of the Standard Model (SM) have been proposed to provide an origin for their mass, with the leading explanation being the Seesaw Mechanism \cite{Minkowski:1977sc, Yanagida:1979as, Glashow:1979nm, GellMann:1980vs,Magg:1980ut,Cheng:1980qt,Lazarides:1980nt,Mohapatra:1980yp,Foot:1988aq,Albright:2003xb}. One class of these models is known as the Type II Seesaw Mechanism, which involves the introduction of a $ SU(2) $ triplet scalar to the SM with Yukawa couplings to the neutrinos. When the neutral component of the triplet Higgs takes its non-zero vacuum expectation value a Majorana mass term is generated for the neutrinos \cite{Magg:1980ut,Cheng:1980qt,Lazarides:1980nt,Mohapatra:1980yp}. The interactions induced by the addition of the triplet Higgs lead to wide-ranging phenomenological implications, which will be probed in future experiments.

An important feature of the Seesaw mechanism is its potential to also explain the origin of the baryon asymmetry of our universe - through a process known as Leptogenesis \cite{Fukugita:1986hr}. The main idea of these scenarios is that an asymmetry initially generated in the leptonic sector during the early universe, is transmitted to the baryonic sector through equilibrium sphaleron processes \cite{Klinkhamer:1984di,Kuzmin:1985mm,Trodden:1998ym,Sugamoto:1982cn}. In approaching the possibility of Leptogenesis in each of the types of Seesaw Mechanism, it is important to consider the testable low-scale predictions. In general, such scenarios in the Type I and Type III Seesaw Mechanisms are typically at very high energy scales, making them difficult to probe. In contrast, the triplet Higgs of the Type II Seesaw Mechanism offers opportunities to naturally connect these high and low-scale dynamics. However, the Type II Seesaw Mechanism is unable to successfully lead to standard thermal Leptogenesis without the inclusion of additional particles; an extra triplet Higgs or a right-handed neutrino \cite{Ma:1998dx}, removing the minimal nature of the model. In recent work, it was found that it is possible to achieve successful Leptogenesis within the minimal Type II Seesaw framework, through the Affleck-Dine Mechanism (ADM) \cite{Affleck:1984fy,Murayama:1992ua, Murayama:1993xu,Barrie:2021mwi,Barrie:2022cub,Han:2022ssz}. Thus, providing an opportunity to test High Scale Leptogenesis in terrestrial experiments. 

Examples of this are the upcoming experimental searches for Lepton Flavour Violation (LFV). The doubly- and singly-charged components of the triplet Higgs lead to LFV decay processes such as $\mu \rightarrow e \gamma$, $\mu \rightarrow e e e$, and $\mu \rightarrow e$ Conversion in Nuclei \cite{Dinh:2012bp,Han:2021nod}. Alongside the induced Majorana neutrino masses which allow for neutrinoless double beta decay. Searches for each of these processes complement each other through their varied dependencies on the triplet Higgs mass, neutrino oscillation parameters, $\mathcal{CP}$ violating phases, mass ordering, and lightest neutrino mass. Future experiments such as Mu3e \cite{Perrevoort:2018okj} and COMET \cite{Moritsu:2022lem}, will allow for vastly improved sensitivities over current constraints, allowing a new probe of the properties of the leptonic sector, and subsequently High Scale Leptogenesis. In this context, it is integral to determine the predicted signatures of the Type II Seesaw Leptogenesis Scenario at these upcoming experiments.

In our initial studies of the Type II Seesaw Leptogenesis scenario, we found that the introduction of the Triplet Higgs of the Type II Seesaw Mechanism to the SM alone can solve each of the open mysteries of the origin of the observed baryon asymmetry, the mechanism for neutrino masses and the inflationary set-up simultaneously \cite{Barrie:2021mwi,Barrie:2022cub}. In this work, we will explore the potential for future Lepton Flavour Violation experiments to test and discover the triplet Higgs of Type II Seesaw Leptogenesis, and the associated dependencies on the properties of the neutrino sector. This paper is structured as follows: In Section \ref{rev_T2SS}, we provide a summary of the Type II Seesaw mechanism, and briefly review how successful Leptogenesis can be achieved by it. Section \ref{LFVs} discusses the Lepton Flavour Violating processes important for triplet Higgs searches, and the associated experimental reach for different neutrino mixing scenarios. The corresponding expected bounds on the allowed parameter space for simultaneous Leptogenesis, neutrino mass generation and inflation are detailed in Section \ref{ex_param}, with dependencies on the neutrino masses and $ \mathcal{CP} $ phases depicted in detail. Finally, in Section \ref{Conc}, we discuss the overall results and the important features for future experimental explorations of Lepton Flavour Violating processes to test Type II Seesaw Leptogenesis.


\section{The Type II Seesaw Mechanism and Leptogenesis}
\label{rev_T2SS}

The Type II Seesaw Mechanism provides a natural framework in which to explain the origin of the neutrino masses. This scenario involves the minimal extension of the SM scalar sector by a $SU(2)_L$ triplet scalar $\Delta$ which carries a hypercharge of $2$. The triplet and SM doublet Higgs' are parameterized as follows,
\begin{eqnarray}
H =\left(
\begin{array}{c}
~~ h^{+} \\
h
\end{array}
\right)
,~~
\Delta =\left(
\begin{array}{cc}
\Delta^+/\sqrt{2} & \Delta^{++} \\
\Delta^0& -\Delta^+/\sqrt{2}
\end{array}
\right)~,
\end{eqnarray}
where $h$ and $\Delta^0$ are the neutral components of $H$  and $\Delta$ respectively. In addition to the neutral component, there exists two charged components of the triplet Higgs, $\Delta^+$ and $\Delta^{++}$. These have important phenomenological implications for terrestrial collider experiments and will be the key to discovering the Type II Seesaw Leptogenesis scenario \cite{Dev:2018sel,BhupalDev:2018tox, Ashanujjaman:2021txz,Chongdar:2021tgm,Dev:2021axj,Mandal:2022zmy,Cheng:2022jyi}. 

The inclusion of the triplet Higgs into the SM leads to new interactions in the Lagrangian that involve the SM Higgs and the left-handed lepton doublets. Firstly, consider the Yukawa interaction that is induced  between the  triplet Higgs $\Delta$ and the left-handed lepton doublets $L_i$ , 
\begin{equation}
{\mathcal L}_{\textrm{Yukawa}}=\mathcal L^{\rm SM}_{\textrm{Yukawa}}-y_{ij}\bar L^c_i \Delta L_j + h.c. 
\label{nu_interaction2}
\end{equation}
Once the neutral component of the triplet Higgs  $\Delta^0$  obtains its non-zero VEV, through this Yukawa interaction a non-zero neutrino mass matrix will be generated. Importantly for our realisation of Leptogenesis,  this interaction also allows us to assign a lepton charge of $Q_L=-2$ to the triplet Higgs. 

New terms in the  Higgs' potential $V(H, \Delta)$ are also induced through the inclusion of the triplet Higgs scalar. These include interactions associated with the $\Delta$ that violate the global lepton number $ U(1)_L $ symmetry. The Higgs' potential has the form,
\begin{eqnarray}
V(H, \Delta) &=& -m_H^2 H^\dagger H + \lambda_H  (H^\dagger H)^2 + m_\Delta^2 {\rm Tr}(\Delta^\dagger \Delta) +  \lambda_1 (H^\dagger H)  {\rm Tr}(\Delta^\dagger \Delta) 
+ \lambda_2 ({\rm Tr}(\Delta^\dagger \Delta))^2  \nonumber \\
&& + \lambda_3 {\rm Tr}(\Delta^\dagger \Delta)^2  + \lambda_4  H^\dagger \Delta \Delta^\dagger H  + \left[\mu( H^T i \sigma^2 \Delta^\dagger H) + \frac{\lambda_5}{M_p} ( H^T i \sigma^2 \Delta^\dagger H) (H^\dagger H)\right. \nonumber \\
&& \left. + \frac{\lambda^\prime_5}{M_p} ( H^T i \sigma^2 \Delta^\dagger H) (\Delta^\dagger \Delta) +h.c.\right] +...~,
\label{full_pot3}
\end{eqnarray}
where the terms within the square brackets $ [...] $ violate the $ U(1)_L $ symmetry. Note that, the VEV obtained by $\Delta^0$ is dependent upon the size of the cubic $\mu$  term. In addition to the cubic term, we have included dimension five operators which are suppressed by the Planck scale, $M_p$. Our Leptogenesis scenario takes place in the early universe when the field values are close to the Planck scale, such that the higher dimensional terms may begin to dominate over the cubic term. These additional terms will not play a direct role in the analysis that follows, as they will be negligible at low energies. It is also important that all of the parameters in this potential ensure vacuum stability up to the Planck Scale. This condition was recently explored in the context of our model in Ref. \cite{Han:2022ssz}.

The VEV of the triplet Higgs can be derived from the potential given above. Taking the limit where the SM Higgs VEV is much smaller than the $\Delta $ mass parameter, i.e. $m_\Delta \gg v_{\textrm{EW}}$, the $\Delta^0$ VEV can be approximated by,
\begin{equation}
v_\Delta \equiv \langle \Delta^0 \rangle \simeq \frac{\mu v_{\textrm{EW}}^2}{ 2 m^2_\Delta}~,
\end{equation}
where the SM Higgs VEV is $v_{\textrm{EW}} = 174$ GeV. 

From the Yukawa interaction term in Eq. (\ref{nu_interaction2}), we obtain the mass matrix of the neutrinos,
\begin{equation}
\left(m_{\nu}\right)_{\ell\ell^{\prime}}\equiv m_{\ell\ell^{\prime}}\simeq2y_{\ell\ell^{\prime}}~
v_{\Delta}~,
\label{yukawa2}
\end{equation}
where the matrix of Yukawa couplings $y_{\ell\ell^\prime}$ is directly related to the PMNS neutrino mixing matrix. An important constraint for the allowed parameter space of our model comes from requiring that the neutrino Yukawa couplings should be smaller than $\mathcal{O}(1)$ to ensure that they remain perturbative up to the Planck scale.

The VEV of $\Delta^0$  has the following allowed range of values,
\begin{equation}
\mathcal{O}(1)\textrm{~GeV}>|\langle\Delta^0\rangle|\gtrsim 0.05\textrm{~eV}~,
\end{equation}
where the lower bound ensures the generation of the observed neutrino masses while also requiring that the Yukawa couplings remain perturbative, and the upper limit is derived from the T-parameter constraints derived from electroweak precision measurements \cite{Kanemura:2012rs}.

Another important constraint on the triplet Higgs comes from the LHC, which has placed a lower limit on the mass parameter of $m_\Delta \gtrsim 800 $ GeV from searches for the associated doubly-charged Higgs \cite{ATLAS:2017xqs}. An important result of this lower bound, is that the masses of the charged and neutral components of the triplet Higgs must be approximately equivalent, $m_{\Delta^{++}} \simeq m_{\Delta^{+}} \simeq m_{\Delta^{0}} \simeq m_\Delta $.


\subsection{Type II Seesaw Leptogenesis}
\label{TIILeptp}

The Type II Seesaw mechanism is known to be unable to successfully lead to standard thermal Leptogenesis, in contrast to the Type I and III Seesaw mechanisms. Thermal Leptogenesis can only be achieved in this mechanism through the inclusion of additional particles, an extra triplet Higgs or a right-handed neutrino \cite{Ma:1998dx}, undoing the minimal nature of the model. However, in recent work, it was found that it is possible to achieve successful Leptogenesis within the minimal Type II Seesaw framework, through the ADM \cite{Affleck:1984fy,Barrie:2021mwi,Barrie:2022cub,Han:2022ssz}. 

The fundamental idea of the ADM is that a non-zero angular motion is generated in the phase of a complex
scalar field $\varphi$. If this scalar is charged under a global $ U(1) $, a corresponding non-zero charge will be produced.
For these dynamics to occur and to lead to successful Baryogenesis, three conditions are required, analogous to the well-known Sakharov conditions for Baryogenesis \cite{Sakharov:1967dj}, namely:
\begin{enumerate}
\item  The scalar is charged under some mixture of the global $U(1)_L$ or  $U(1)_B$ symmetries,
\item There exists a term in the Lagrangian that violates $U(1)_L$ or  $U(1)_B$,
\item The scalar has a displaced vacuum value during the early universe.
\end{enumerate}
These requirements are easily seen by considering the charge number density of the complex scalar in the polar coordinates,
\begin{equation*}
n_\varphi = j^0  = 2 Q \textrm{Im}[\varphi^\dagger \dot \varphi]= Q\varphi_r^2\dot{\theta}~,
\end{equation*}
where $ \varphi = \frac{1}{\sqrt{2}}\varphi_r e^{i\theta} $ and $ Q $ is its global $ U(1) $ charge.

Interestingly, the triplet Higgs of the Type II Seesaw mechanism immediately fulfils the first two of these conditions. Namely, the triplet Higgs is assigned a $U(1)_L$ charge through its Yukawa couplings to the leptons, and it inherently has $U(1)_L$ breaking interactions in the Lagrangian through its couplings to the SM Higgs. In our analysis, we have considered the scenario where the dimension five lepton violating terms provide the dominant contribution to the lepton asymmetry generation during inflation, to allow for numerical and analytical treatment. This is not possible in the case of a dominant $ \mu $ coupling due to its behaviour during the oscillation phase after inflation, which if significant, requires a detailed analysis of the preheating and reheating phase to determine the lepton asymmetry predictions \cite{Ema:2016dny, DeCross:2015uza, DeCross:2016cbs, DeCross:2016fdz,Sfakianakis:2018lzf,Ema:2021xhq}. For details of the $ \mu $ dynamics see our recent work in Ref. \cite{Barrie:2022cub}.

To satisfy condition 3, we consider that the triplet Higgs is a component of the inflaton in combination with the SM Higgs \cite{Brout:1977ix, Sato:1980yn, Guth:1980zm, Linde:1981mu, Albrecht:1982mp,Starobinsky:1980te,Whitt:1984pd,Jakubiec:1988ef,Maeda:1988ab,Barrow:1988xh,Faulkner:2006ub,Bezrukov:2007ep,Bezrukov:2008ut,GarciaBellido:2008ab,Barbon:2009ya,Barvinsky:2009fy,Bezrukov:2009db,Giudice:2010ka,Bezrukov:2010jz,Burgess:2010zq,Bezrukov:2011gp,Hertzberg:2013mba, Lozanov:2014zfa, Yamada:2015xyr, Bamba:2016vjs,Bamba:2018bwl, Cline:2019fxx,Barrie:2020hiu, Lin:2020lmr, Kawasaki:2020xyf, Kusenko:2014lra, Wu:2019ohx, Charng:2008ke, Ferreira:2017ynu, Rodrigues:2020dod, Lee:2020yaj, Enomoto:2020lpf, Mohapatra:2021aig, Mohapatra:2022ngo}. To do this, we require the introduction of non-minimal couplings to gravity for both the SM and triplet Higgs' that are of the following form,
\begin{eqnarray}
( \xi_H |h|^2 + \xi_\Delta |\Delta^0|^2)R=\left(\frac{1}{2}\xi_H \rho^2_H+\frac{1}{2} \xi_\Delta \rho^2_\Delta\right)R~,
\end{eqnarray}
where $ R $ is the Ricci scalar, we have used the following polar coordinate parametrization  $h \equiv \frac{1}{\sqrt{2}} \rho_{H} e^{i\eta}$,  $\Delta^0 \equiv \frac{1}{\sqrt{2}}\rho_{\Delta} e^{i\theta}$, and have utilised the unitary gauge. Interestingly, this set-up exhibits a unique inflationary trajectory given by the following ratio of the radial components of the two Higgs' \cite{Lebedev:2011aq,Lee:2018esk,Choi:2019osi},
\begin{equation}
\frac{\rho_H}{\rho_\Delta} = \tan\alpha= \sqrt{\frac{ 
2 \lambda_\Delta \xi_H - \lambda_{h\Delta} \xi_\Delta }{2 \lambda_H \xi_\Delta - \lambda_{h\Delta} \xi_H} }  ~,
\label{traj}
\end{equation}
where $ \lambda_\Delta = \lambda_2+\lambda_3~, $ and  ~$ \lambda_{H\Delta} = \lambda_1+\lambda_4 $ from Eq. (\ref{full_pot3}). The requirement imposed on these couplings from Cosmic Microwave Background (CMB) observables  is given by \cite{Planck:2018jri},
\begin{equation}
\frac{\lambda}{\xi^2}=\frac{\lambda_H \sin^4 \alpha + \lambda_{H\Delta} \sin^2 \alpha \cos^2 \alpha + \lambda_{\Delta} \cos^4 \alpha}{(\xi_H \sin^2 \alpha + \xi_\Delta \cos^2 \alpha)^2}\simeq 5\cdot 10^{-10} ~.
\label{CMB_con}
\end{equation}

Note that, the exact dynamics of the inflationary epoch and lepton number generation are not integral to the analysis that we undertake here, which concerns low energy phenomena. However, the constraints applied by these experiments are important to understanding the running of the scalar couplings up to the Planck Scale. These couplings determine the dominant component of the inflaton and subsequently which dimension five operator leads to Leptogenesis, while establishing the inflationary trajectory and the required non-minimal couplings to gravity for consistency with CMB observables - as described in Eq. (\ref{traj}) and (\ref{CMB_con}). For details of the full dynamics of our model in the early universe, see our recent work in Ref. \cite{Barrie:2022cub}.

Thus, we arrive at a simple framework that successfully explains the observed baryon asymmetry, the origin of the non-zero neutrino masses, and the set-up of the inflationary epoch. 
Note that, this usage of the ADM also possesses additional advantages over other realisations - not requiring Supersymmetry, providing a natural path for transferring the lepton asymmetry generated in the scalar sector to the SM fermions, and providing testable low energy phenomenological predictions associated directly with the inflaton field.


\subsection*{Parameter Constraints from Lepton Number Washout Effects} 

Given that the inflaton is made up of scalars that couple to gauge bosons, we expect a large reheating temperature, $ T_{\textrm{rh}}\sim 10^{13-14} $ GeV. Subsequently, the triplet Higgs will be quickly thermalised at the end of reheating, and it is necessary to consider possible lepton asymmetry washout processes. These can provide important constraints on the allowed triplet Higgs parameters for successful  Leptogenesis to occur.  The main processes that are necessary to consider are $LL\leftrightarrow\Delta$ and $ HH\leftrightarrow\Delta $. If they co-exist, the lepton number generated during inflation will be rapidly washed out. However, to transmit the asymmetry from the scalar sector to the leptonic sector, the process $LL\leftrightarrow\Delta$ must be efficient while  $HH\leftrightarrow\Delta $ is out of equilibrium. It is easy to ensure that the first process is in equilibrium for parameter values in the range $m_\Delta < 10^8$ GeV and $y> 10^{-5}$. To prevent the second process from acting efficiently in the early universe, we have the following requirement,
\begin{eqnarray}
\Gamma_{ID}(HH\leftrightarrow \Delta)|_{T=m_\Delta}<H|_{T=m_\Delta}~,
\label{Decay2}
\end{eqnarray}
where $\Gamma_{ID}(HH\leftrightarrow \Delta)|_{T=m_\Delta} \approx \Gamma_{D}(\Delta \rightarrow HH) \simeq \frac{\mu^2}{32\pi m_\Delta}$ and $|v_\Delta| \simeq  \frac{\mu v_{\textrm{EW}}^2}{ 2 m^2_\Delta}$. From Eq. (\ref{Decay2}) we derive the following constraint,
\begin{eqnarray}
v_\Delta \lesssim 10 \textrm{~keV} \left( \frac{m_\Delta}{1 \textrm{~TeV} }\right)^{-1/2} ~,
\label{con_vev1}
\end{eqnarray}
which corresponds to a limit of $v_\Delta \lesssim 10$ keV for $m_\Delta \gtrsim 1$ TeV, which we can also express as a lower bound on the largest neutrino Yukawa coupling,
\begin{eqnarray}
y\gtrsim 2.2 \cdot 10^{-6} \sqrt{ \frac{m_\Delta}{800 \textrm{~GeV} }} ~,
\label{con_y1}
\end{eqnarray}
such that the largest neutrino mass is given approximately by the mass difference $ \sqrt{\Delta m_{31}^2} $. Importantly, the LFV experiments provide $ m_\Delta $ dependent upper limits on the coupling $ y $.

These two equivalent constraints on the vacuum expectation value (Eq. (\ref{con_vev1})) and largest neutrino Yukawa coupling (Eq. (\ref{con_y1})) have important ramifications for triplet Higgs searches at collider experiments. Depending on the vacuum expectation value and triplet Higgs Yukawa couplings, the dominant decay channel of the triplet Higgs is either gauge bosons or leptons, for small mass splittings between the neutral and charged components of the triplet Higgs. This is because larger vacuum expectation values require smaller Yukawa couplings to the leptons to not produce neutrino masses that are too large. In turn, this suppresses the triplet Higgs decays into leptons, leading to decays into gauge bosons being its dominant decay process. In order to avoid the washout of our lepton asymmetry after reheating, we require $v_\Delta<10$ keV. This bound requires that the triplet Higgs dominantly decays into leptons, for small triplet Higgs mass splittings, providing a unique prediction for our Leptogenesis mechanism \cite{Melfo:2011nx}. Thus, providing a complementary test to the expected LFV signatures discussed below.


\subsection*{Requirement of Vacuum Stability for the Inflationary Setting} 

The Type II Seesaw Leptogenesis scenario involves an inflationary setting induced by a combination of the SM Higgs and triplet Higgs, which have non-minimal couplings to gravity. A requirement of this set-up is that the Higgs' vacuum remains stable up to the Planck Scale. This is not the case for the current best fit parameters of the SM alone, which is found to be metastable, exhibiting an instability scale at around $ \sim 10^{10} $ GeV \cite{EliasMiro:2011aa,Degrassi:2012ry,Lebedev:2012sy,Salvio:2013rja,Branchina:2014usa,Bezrukov:2014ipa}. Note, this result is sensitive to the top pole mass, which is currently not determined with high precision. The addition of the Triplet Higgs to the SM, and its corresponding couplings, influence the running of the SM couplings as well as alter the requirements for vacuum stability. In particular, the triplet Higgs Yukawa couplings to the lepton doublets play an important role in the running of these couplings and influence the parameters at high scales.  The conditions for vacuum stability  in the Type II Seesaw mechanism up to the Planck Scale has been explored in Ref. \cite{Bonilla:2015eha, Moultaka:2020dmb,Han:2022ssz}. An example constraint, is that the mass splitting between the components of the triplet Higgs are generally required to be less than 10 GeV for $m_\Delta>1$ TeV \cite{Bonilla:2015eha}. This is particularly important for determining that the dominant decay of the triplet is leptonic in our scenario, as discussed above. However, it should be noted that these analyses did not include the running of the non-minimal couplings and the corresponding CMB requirements on the couplings in Eq. (\ref{CMB_con}).  

The non-minimal couplings and inflationary trajectory itself are dependent upon the values of the various couplings and mass parameters in the scenario, so it is important to understand all possible constraints on the allowed parameter space. The inflationary  trajectory is particularly sensitive to the quartic couplings of the SM $ \lambda_H $ and Triplet Higgs' $ \lambda_\Delta $, the portal coupling $ \lambda_{H\Delta} $, and the non-minimal couplings - as shown in Eq. (\ref{traj}) and (\ref{CMB_con}). Additionally, these couplings establish the component of the inflaton that dominates the inflationary dynamics, which subsequently determines which dimension five lepton violating interaction in Eq. (\ref{full_pot3}) is responsible for Leptogenesis. If Leptogenesis is able to be induced by the $ \mu $ coupling, the differing operator dependence on the $ h $ and $ \Delta^0 $ components will also make it sensitive to which field is the dominant component of the inflaton. LFV experiments provide constraints on the phase space of the $ m_\Delta  $ and $ y_{\ell\ell^\prime}  $ parameters, and thus, will be a necessary complementary probe to the inflationary observables and the realisation of successful Leptogenesis in our scenario.


\section{Lepton Flavour Violation Processes and Expected Signatures}
\label{LFVs}

The Type II Seesaw Mechanism can be probed through precision tests of rare processes in the leptonic sector. These are associated with the new couplings introduced between the triplet Higgs and the leptons through its charged components, as well as the generation of a Majorana mass term for the neutrino. Important LFV decays induced by the triplet Higgs are the $\mu \rightarrow e \gamma$, $\mu \rightarrow e e e$, and $\mu \rightarrow e$ Conversion in Nuclei processes. The branching ratios of each of these decays have different dependencies on the neutrino oscillation properties and triplet Higgs parameters. Thus, it is integral to search experimentally for the predicted signatures in each of these decay processes. In this section, we will survey the sensitivities of current and future experiments searching for evidence of LFV, and how these limits translate to constraints on the triplet Higgs of our model, and subsequently the Leptogenesis mechanism and inflationary scenario. 

In our analysis below, we follow the notation utilised in Ref. \cite{Dinh:2012bp}.  The triplet Higgs Yukawa couplings in Eq. (\ref{yukawa2}) are given by, 
\begin{equation}
y_{\ell\ell^\prime}= \frac{1}{2v_\Delta}\left(U^*{\rm diag}(m_1,m_2,m_3)U^\dagger\right)_{\ell\ell^\prime}~,
\label{yukawa1}
\end{equation}
where $ m_i $ are the masses of the neutrino mass eigenstates, and $U$ is the unitary PMNS neutrino mixing matrix. In the standard parametrisation it takes the form,
\begin{equation}
U = V(\theta_{12},\theta_{23},\theta_{13},\delta)Q(\alpha_{21},\alpha_{31})~,
\label{U_PMNS}
\end{equation}
where
\begin{equation}
V = \left(
\begin{array}{ccc}
1 & 0 & 0 \\
0 & c_{23} & s_{23} \\
0 & -s_{23} & c_{23} \\
\end{array}\right)\left(
\begin{array}{ccc}
c_{13} & 0 & s_{13}e^{-i\delta} \\
0 & 1 & 0 \\
-s_{13}e^{i\delta} & 0 & c_{13} \\
\end{array}
\right)\left(
\begin{array}{ccc}
c_{12} & s_{12} & 0 \\
-s_{12} & c_{12} & 0 \\
0 & 0 & 1 \\
\end{array}
\right)~,
\label{V_PMNS}
\end{equation}
and 
$c_{ij}=\cos\theta_{ij}$, $s_{ij}=\sin\theta_{ij}$, $\delta$ is the Dirac $\mathcal{CP}$ phase, and the matrix $Q$ contains the two Majorana $\mathcal{CP}$ phases, $ \alpha_{21} $ and $ \alpha_{31} $,
\begin{equation}
Q =\left(\begin{array}{ccc}
1 & 0 & 0 \\
0 & e^{i\alpha_{21}/2} & 0 \\
0 & 0 & e^{i\alpha_{31}/2}) \\
\end{array}\right) ~.
\label{Q_maj}
\end{equation}

The neutrino oscillation parameters and mass differences used for both the  Normal Ordering (NO) and Inverted Ordering (IO) scenarios are presented in Table  \ref{nufit_params}, which correspond to the best fit parameters (BFP) for the current neutrino mixing data including Super-Kamiokande atmospheric neutrino data. The $ 1\sigma $ values are included in the Table, but we take the central values in our analysis.

\begin{table}[h]
\caption{Best fit parameters (BFP) for the neutrino mixing angles, $\mathcal{CP}$ phase, and mass differences - for both the Normal Ordering (NO) and Inverted Ordering (IO) scenarios. These are the global fit values determined by Nufit \cite{Esteban:2020cvm}, which include the Super-Kamiokande atmospheric neutrino data, and which we use for our analysis.}
	\begin{center}
		\begin{tabular}{|c|c c|}
			\hline 
			BFP $\pm 1\sigma$ &  NO & IO  \\
			\hline
			$ \sin^2 \theta_{12} $ ~~&~~ $ 0.304^{+0.012}_{-0.012} $~~ &~~ $ 0.304^{+0.013}_{-0.012} $~~ \\
			$ \sin^2 \theta_{23} $ ~~&~~ $ 0.450^{+0.019}_{-0.016} $ ~~&~~ $ 0.570^{+0.016}_{-0.022} $~~ \\
			$ \sin^2 \theta_{13} $ ~~&~~ $ 0.02246^{+0.00062}_{-0.00062} $~~&~~ $ 0.02241^{+0.00074}_{-0.00062} $~~  \\
			$  \delta_{\mathcal{CP}} $ ~~&~~ $ 230^{+36}_{-25} $&~~ $ 278^{+22}_{-30} $~~ \\
			$  \frac{\Delta m^2_{21} }{10^-5~\textrm{eV}^2}$ ~~&~~ $ 7.42^{+0.21}_{-0.20} $~~&~~ $ 7.42^{+0.021}_{-0.020} $~~ \\
			$  \frac{\Delta m^2_{31} }{10^-3~\textrm{eV}^2} $ ~~&~~ $ +2.510^{+0.027}_{-0.027} $ ~~&~~ $ -2.490^{+0.026}_{-0.028} $~~
			\\\hline
		\end{tabular}
		\label{nufit_params}
	\end{center}
\end{table}

The existence of the physical Majorana phases,  $ \alpha_{21} $ and $ \alpha_{31} $, in the PMNS matrix was first noted in Ref. \cite{Bilenky:1980cx}. There are no current measurements of the Majorana phases from the neutrino oscillation experiments. This is because the flavour neutrino oscillations are not sensitive to the Majorana phases, and thus to whether the massive neutrinos are of Dirac or Majorana nature, for both oscillations in the vacuum \cite{Bilenky:1980cx}, and in matter \cite{Langacker:1986jv}.

Before beginning the analysis, we note some important assumptions and features of our model. For the triplet Higgs, these LFV processes are dominantly mediated by the double charged component. In our analysis, we assume that the mass differences between each of the components of the triplets are small, i.e. $m_\Delta^0\simeq m_\Delta^+\simeq m_\Delta^{++}=m_\Delta$ . Additionally, the Leptogenesis mechanism we consider here is independent of the leptonic $\mathcal{CP}$ phases, rather  $\mathcal{CP}$ is spontaneously broken during the early stages of the universe. This means that we must consider all possibilities for the neutrino $\mathcal{CP}$ phases in our analysis. The upper limits we use for the lightest neutrino masses are from the current cosmological bound on the sum of the neutrino masses - $\sum m_\nu< 0.12$ \cite{Aghanim:2018eyx}. This corresponds to approximate upper limits of $ m_1<0.03 $ eV for NO, and $ m_3<0.015 $ eV for IO. Additionally, given the current uncertainty in the measured $\delta_{\mathcal{CP}}$ phase, we allow it to vary while also providing results for the best fit value for each neutrino mass ordering.


\subsection{Current $\mu \rightarrow e\gamma$ Constraints}

The first LFV decay process we consider is $\mu \rightarrow e \gamma$, which is generated at the one-loop level by the doubly- and singly-charged components of the triplet Higgs. The branching ratio for this process is given by \cite{Dinh:2012bp},
\begin{equation}
\textrm{BR}(\mu \rightarrow e\gamma) \simeq\frac{\alpha_{\rm em}}{192\pi}\frac{\left|\left(y^{\dagger}y\right)_{e \mu}\right|^{2}}{G_{F}^{2}}\left(\frac{1}{m^2_{\Delta^{+}}} +\frac{8}{m^2_{\Delta^{++}}}\right)^{2}\simeq\frac{81\alpha_{\rm em}}{192\pi}\frac{\left|\left(y^{\dagger}y\right)_{e \mu}\right|^{2}}{G_{F}^{2} m^4_{\Delta^{++}}}~,
\label{mueg1}
\end{equation}
where $ \alpha_{\textrm{em}} $ is the fine structure constant, and $ G_F $ is the Fermi constant.

At present, the strongest constraints on this process have been determined by the MEG collaboration, with an upper bound on the branching ratio of \cite{MEG:2016leq},
\begin{equation}
\textrm{BR}(\mu \rightarrow e\gamma) < 4.2 \cdot 10^{-13}~.
\end{equation}

From this limit, we can derive a bound on the Yukawa couplings in the case of $ m_\Delta^+\simeq m_\Delta^{++}=m_\Delta$ ,
\begin{equation}
\left|\left(y^{\dagger}y\right)_{e \mu}\right|<1.6 \cdot 10^{-4}\left(\frac{m_{\Delta}}{800 \textrm{~GeV}}\right)^2~,
\label{con_mueg1}
\end{equation}
from which we derive the following $ m_\Delta $ dependent lower bound on the triplet Higgs parameter $ \mu $,
\begin{equation}
\mu> 1.7 \cdot 10^{-6} \textrm{~GeV} \frac{\sqrt{\left|\left(m^{\dagger}m\right)_{e\mu}\right|}}{ 1 \textrm{~eV}}\frac{m_{\Delta}}{800 \textrm{~GeV}}~,
\label{con_mueg2}
\end{equation}
where we have used the relation $|v_\Delta|=\frac{\mu v_{\textrm{EW}}^2}{2 m_\Delta^2}$ .

To calculate the mass parameter we use the fact that the term $ \left|\left(m^{\dagger}m\right)_{e\mu}\right| $ can be simply expressed in terms of the neutrino oscillation and mass difference parameters as follows,
\begin{equation}
\left|\left(m^{\dagger}m\right)_{e\mu}\right|=4v^2_{\Delta}\left |\left(h^{\dagger}h\right)_{e\mu}\right| =
\left | U_{e2}U^{\dagger}_{2\mu} \Delta m^2_{21} +
U_{e3}U^{\dagger}_{3\mu} \Delta m^2_{31} \right|~.
\label{mass_param}
\end{equation}

\begin{figure}[h]
\centering
\includegraphics[width=0.6375\columnwidth]{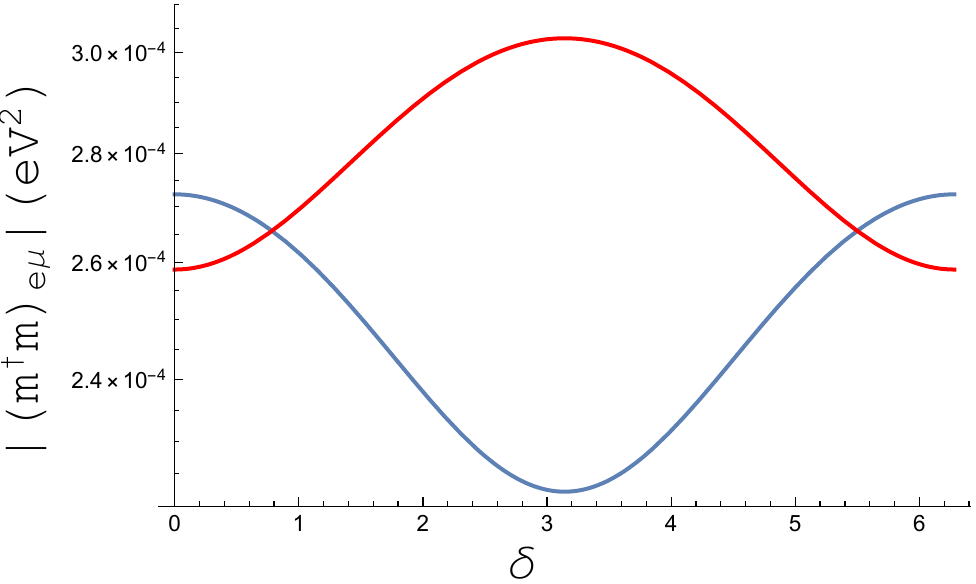}
\caption{The $ \left|\left(m^{\dagger}m\right)_{e\mu}\right| $ parameter, constrained by the $ \mu\rightarrow e\gamma $ process at the MEG experiment \cite{MEG:2016leq}, as a function of the $\delta_\mathcal{CP}$ phase for the NO (Blue) and IO (Red) scenarios. All other parameters are fixed by the corresponding best fit parameters in Table \ref{nufit_params}.}
\label{mu_eg} 
\end{figure}

Thus, the $ \mu\rightarrow e\gamma $ branching ratio for the triplet Higgs is independent of the lightest neutrino mass and Majorana neutrino phases, varying only through the $\delta_{\mathcal{CP}}$ phase once the mixing angles and mass differences are fixed. The best fit neutrino oscillation and mass difference parameters given in Table \ref{nufit_params} are taken, and the relation in Eq. (\ref{mass_param})  depicted in Figure \ref{mu_eg} with variation with respect to $ \delta_{\mathcal{CP}} $ and the neutrino mass ordering. It can be seen that the parameter value is relatively weakly dependent on the $ \delta_{\mathcal{CP}} $ phase and the type of ordering. For illustrative purposes, if we take the approximate value where the two curves intersect, we find that,
\begin{equation}
\left|\left(m^{\dagger}m\right)_{e\mu}\right|\sim 2.65\cdot 10^{-4} \textrm{~eV}^2 ~,
\end{equation}
which when substituted into Eq. (\ref{con_mueg1}) gives the following approximate maximum reach on the mass parameter of the triplet Higgs, $ m_\Delta \lesssim 2\cdot 10^4 $ GeV, when taking the maximal neutrino Yukawa coupling to be $y \sim 1 $. This also gives the following lower bound on the cubic coupling $ \mu $,
\begin{equation}
\mu> 2.8 \cdot 10^{-8} \textrm{~GeV}~\frac{m_{\Delta}}{800 \textrm{~GeV}}~.
\label{con_mueg3}
\end{equation}

Importantly, this constraint can be reinterpreted as an approximate $ m_{\Delta} $ dependent upper bound on the largest neutrino Yukawa coupling $ y $,
\begin{equation}
y< 0.038 ~\frac{m_{\Delta}}{800 \textrm{~GeV}}~, 
\label{y_con_mueg}
\end{equation}
where $ y=1 $ gives the maximum sensitivity to the $ m_{\Delta} $ parameter before the Yukawa couplings become non-perturbative. This bound is an effective lower bound on the vacuum expectation value of the triplet, assuming the limit provided on $ \mu $ and the largest neutrino mass is given approximately by the mass difference $ \sqrt{\Delta m_{31}^2} $. Note the lower bound on the maximal neutrino Yukawa coupling provided by the lepton asymmetry washout constraint in Eq. (\ref{con_y1}).

As will be shown in the next Section, searches for the $ \mu\rightarrow e\gamma $ decay process at MEG currently provide the best bounds on the triplet Higgs properties in the case of NO for various parameter sets, except for when the neutrino masses approach the Quasi-Degenerate regime. However, this is not the case in the IO scenario, where constraints on $\mu \rightarrow 3e$ from the SINDRUM experiment tend to dominate \cite{SINDRUM:1987nra}. The $ \mu\rightarrow e\gamma $ decay process also exhibits a small dependence on $\delta_{\mathcal{CP}}$ compared to the sensitivity to the $ \mathcal{CP} $ phases of the other LFV processes considered below.


\subsection{Current $\mu \rightarrow 3e$ Constraints and Future Sensitivities}

The doubly-charged component of the triplet Higgs leads to tree level $\mu\rightarrow 3e$ decay processes, which provide an important test of the triplet Higgs couplings. The branching ratio of this process has the following form,
\begin{equation}
\textrm{BR}(\mu\rightarrow 3e) =\frac{1}{G_F^2}\frac{|(y^\dagger)_{ee}(y)_{\mu e}|^2}{m^4_{\Delta^{++}}} =
\frac{1}{G_F^2 m^4_{\Delta}}
\frac{|m^*_{ee}m_{\mu e}|^2}{16v^4_{\Delta}}~.
\label{IIBRmu3e}
\end{equation}

The current best upper bound on the $\mu \rightarrow 3e$ branching ratio is derived from the results of the SINDRUM experiment \cite{SINDRUM:1987nra},
\begin{eqnarray}
\textrm{BR}(\mu^+\rightarrow e^+ e^- e^+) < 10^{-12}~,
\end{eqnarray}
from this limit, we can determine the present bound on the triplet Higgs Yukawa couplings,
\begin{equation}
|(y^\dagger)_{ee}(y)_{\mu e}| < 7.5 \cdot 10^{-6}\left(\frac{m_{\Delta}}{800 \textrm{~GeV}}\right)^{2}~,
\end{equation}
and correspondingly a limit on the cubic coupling $ \mu $,
\begin{equation}
\mu > 7.7 \cdot 10^{-6} \textrm{~GeV}\frac{\sqrt{|m^*_{ee}m_{\mu e}|}}{1 \textrm{~eV}} \frac{m_{\Delta}}{800 \textrm{~GeV}}~.
\end{equation}

The upcoming experiment known as Mu3e promises to deliver significantly improved sensitivity to the $\mu \rightarrow 3e$ process, potentially probing the branching ratio by an additional four orders of magnitude \cite{Perrevoort:2018ttp}, that is,
\begin{eqnarray}
\textrm{BR}_{\textrm{Mu3e}}(\mu\rightarrow 3e) < 10^{-16}~,
\end{eqnarray}
which will provide the corresponding limit on the Yukawa couplings,
\begin{equation}
|(y^\dagger)_{ee}(y)_{\mu e}| < 7.5 \cdot 10^{-8}\left(\frac{m_{\Delta}}{800 \textrm{~GeV}}\right)^{2}~,
\end{equation}
and on the cubic coupling $ \mu $,
\begin{equation}
\mu > 7.7 \cdot 10^{-5} \textrm{~GeV}\frac{\sqrt{|m^*_{ee}m_{\mu e}|}}{1 \textrm{~eV}} \frac{m_{\Delta}}{800 \textrm{~GeV}}~,
\end{equation}
providing an order of magnitude improvement to the sensitivity to $ \mu $.

Unlike the $ \mu\rightarrow e\gamma $ process considered above, the branching ratio of the $ \mu\rightarrow 3e $ process is highly sensitive to the Majorana phases and type of neutrino mass ordering. Thus, below we will survey the expected reach of the Mu3e experiment under variations of these two parameters, along with the lightest neutrino mass in each scenario. The best fit parameters given in Table \ref{nufit_params} will be fixed throughout, apart from the $ \delta_{\mathcal{CP}} $ phase for which current constraints still allow for significant variation.


\subsubsection*{Sensitivities for NO: $\mu \rightarrow 3e$}

\begin{figure}[h]
\centering
\includegraphics[width=0.65\textwidth]{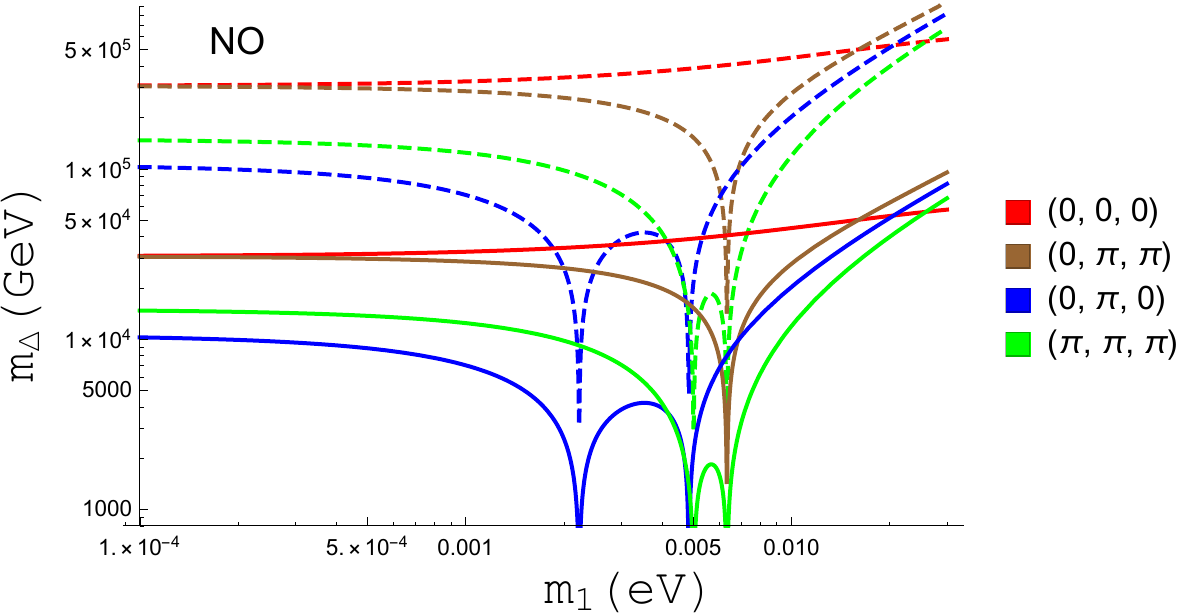}
\caption{Comparison of the maximum expected $ m_\Delta $ probed by the  upcoming $ \mu\rightarrow 3e $ experiment, Mu3e \cite{Perrevoort:2018ttp} (Dashed), and the current constraints from the SINDRUM experiment \cite{SINDRUM:1987nra} (Solid) for different sets of neutrino $\mathcal{CP}$ phases ($ \delta_{\mathcal{CP}} $, $ \alpha_{21} $, $ \alpha_{31} $) and varying lightest neutrino mass, in the NO scenario. All other parameters are given by the best fit parameters in Table \ref{nufit_params}.}
\label{NO_mu_3e}
\end{figure}

Firstly, we will determine the current and projected future constraints in the NO scenario. To do this we consider the  $\mathcal{CP}$ phase parameter sets that maximise and minimise the $ |m^*_{ee}m_{\mu e}| $ parameter. Taking the small $ m_1 $ limit, the $| m_{ee} |$ component is found to be maximised when the $\mathcal{CP}$ phases satisfy $\alpha_{21}-\alpha_{31}+2\delta = \pi$, and minimised when this relation is equal to 0. This corresponds to the following range for the best fit values, $1.43\cdot 10^{-3}~{\rm eV} < |m_{ee}| < 3.69\cdot 10^{-3}$ eV. This is not the case once the lightest neutrino mass becomes greater than $ 10^{-3} $ eV, for which the value can become zero for certain $\mathcal{CP}$ phase combinations and masses, as first noticed in Ref. \cite{Dinh:2012bp}. These suppression scenarios occur when the following relations are satisfied:  $\alpha_{21}=\pi$, and $(\alpha_{31}-2\delta) = 0 $ or $\pi$. This is illustrated in Figure \ref{NO_mu_3e}, with an example zero at $ m_1\simeq 0.00634 $ eV for ($ \delta_{\mathcal{CP}} $, $ \alpha_{21} $, $ \alpha_{31} $)=($ 0 $, $\pi$, $\pi$) which is a result of this behaviour. Such cases are important for our investigations, as they point to the necessity for and complementary nature of searches for each of the LFV decay processes.

The appearance of zeros is also caused by the variation of the $ |m_{\mu e}| $ component with the $ \mathcal{CP} $ phases, which occurs when the $\mathcal{CP} $ phases satisfy ($ 0 $, $\pi$, $ 0 $) and ($ \pi $, $\pi$, $\pi$) \cite{Dinh:2012bp}. This accounts for the additional zeros seen for these parameter sets compared to the ($ 0 $, $\pi$, $\pi$) case, as depicted in Figure \ref{NO_mu_3e}. On the other hand, this component is maximised when $\alpha_{31} - \alpha_{21} = \delta$, and $\delta = \pi$, giving a value of $|m_{\mu e}| < 8.1\cdot 10^{-3}$ eV.

Thus, we can now depict the current and expected future maximal constraints on the triplet Higgs mass parameter for different $ \mathcal{CP} $ phase parameter sets and varying $ m_1 $; see Figure \ref{NO_mu_3e}. The maximum sensitivity is achieved when  $\alpha_{31} - \alpha_{21} = \delta = 0$ is satisfied for small $ m_1 $, with the appearance of a zero as described above for the ($ 0 $, $\pi$, $\pi$) scenario. By fixing this  $ \mathcal{CP} $ phase relation we can arrive at a simple expression for the maximum value of the $ |m^*_{ee}m_{\mu e}| $ parameter in the limit $ m_1\rightarrow 0 $,
\begin{align}
|m^*_{ee}m_{\mu e}|=\left |\left(m_2 s^2_{12}c^2_{13} + m_3 s^2_{13}\right)c_{13}\left( m_2s_{12}(c_{12}c_{23} - s_{12}s_{23}s_{13})+m_3 s_{23}s_{13}\right)\right|~,
\label{max_NO}
\end{align}
which is $ \sim 2.81\cdot 10^{-5}~\textrm{eV}^2 $ for the best fit parameters in Table \ref{nufit_params}. Using this result, we find the current and future expected lower bounds on the cubic coupling $ \mu $ to be,

\begin{equation}
\mu > 4 \cdot 10^{-8} \textrm{~GeV}\frac{m_{\Delta}}{800 \textrm{~GeV}}~,~~\textrm{and}~~
\mu > 4 \cdot 10^{-7} \textrm{~GeV}\frac{m_{\Delta}}{800 \textrm{~GeV}}~,
\end{equation}
respectively. Interestingly, both of these maximal results are greater than the current best
limit from the $ \mu\rightarrow e\gamma $ constraints found in Eq. (\ref{con_mueg3}). We can then translate these bounds into constraints on the largest neutrino Yukawa coupling,
\begin{equation}
y< 0.026 \frac{m_{\Delta}}{800 \textrm{~GeV}}~,~~\textrm{and}~~y< 0.0026\frac{m_{\Delta}}{800 \textrm{~GeV}}~. 
\label{y_con_mu3eNO}
\end{equation}


\subsubsection*{Sensitivities for IO: $\mu \rightarrow 3e$}

Now we move on to the IO case, which in contrast to the NO case does not exhibit zeros. This can be first seen in the $ m_{ee} $ component which is found to take the following minimum value, $|m_{ee}|\geq\sqrt{|\Delta m^2_{31}| + m^2_3}\cos2\theta_{12}\geq 1.96\cdot 10^{-2}$ eV, with the lower limit corresponding to $ m_3=0 $. Considering the $|m_{\mu e}|$ component, it is observed that its maximal value is achieved when the $ \mathcal{CP} $ phases satisfy $\delta = 0$ and
$\alpha_{21} = \pi$, and the lightest neutrino mass $ m_3\rightarrow 0 $. In this case, it takes the value $|m_{\mu e}|= \sqrt {|\Delta m^2_{31}|} c_{13}
(c_{23}\sin 2\theta_{12} + s_{23}s_{13}\cos2\theta_{12})= 3.19\cdot 10^{-2}$ eV, for the best fit parameters. On the other hand,  $|m_{\mu e}|$ experiences significant suppression when $\delta_{\mathcal{CP}} \sim \pi/2$ and the following relation is satisfied,
\begin{equation}
c_{23} c_{12} s_{12} \sin\alpha_{21} \simeq\left (c^2_{12} + s^2_{12}\cos\alpha_{21}\right) s_{23} s_{13}~,
\label{IO_sup}
\end{equation}
from which we derive the value $\alpha_{21}\simeq 0.375 $ for the best fit parameters in Table \ref{nufit_params}. Note, in this case, the suppression is still less severe than the zeros seen in the NO scenario.

\begin{figure}[h]
\centering
\includegraphics[width=0.65\textwidth]{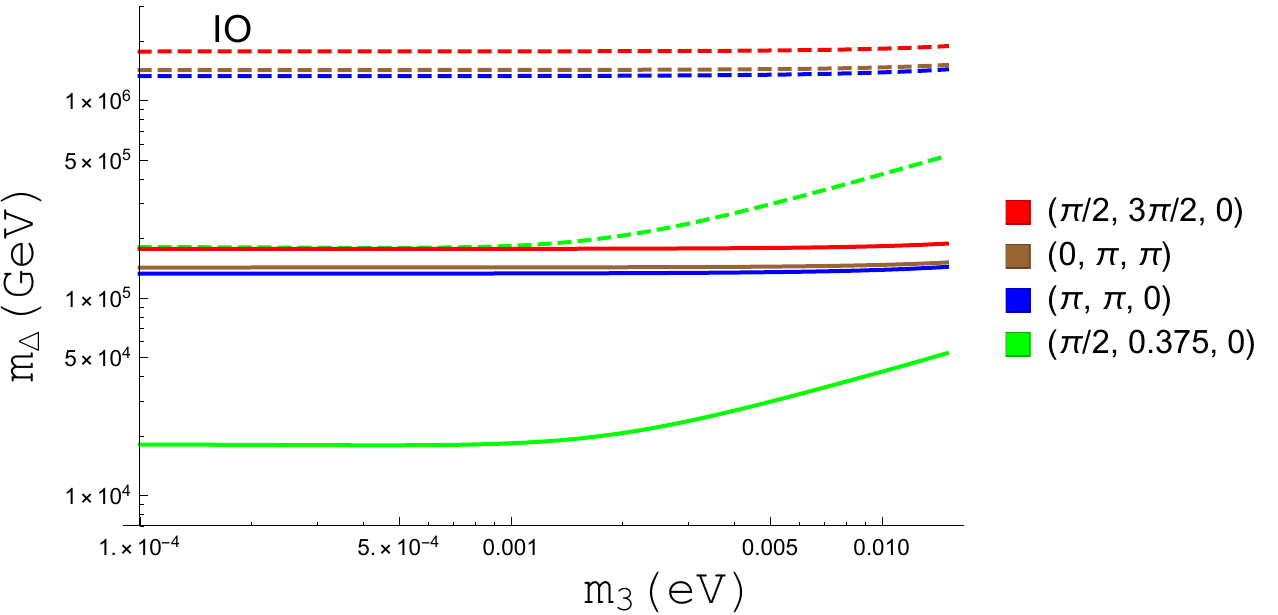}
\caption{Comparison of the maximum expected $ m_\Delta $ probed by the  upcoming $ \mu\rightarrow 3e $ experiment, Mu3e \cite{Perrevoort:2018ttp} (Dashed), and the current constraints from the SINDRUM experiment \cite{SINDRUM:1987nra} (Solid) for different sets of neutrino $\mathcal{CP}$ phases ($ \delta_{\mathcal{CP}} $, $ \alpha_{21} $, $ \alpha_{31} $) and varying lightest neutrino mass, in the IO scenario. All other parameters are given by the best fit parameters in Table \ref{nufit_params}.}
\label{IO_mu_3e}
\end{figure}

Putting these components together, we find that the maximum value for the $|m_{ee}m_{\mu e}|$ parameter for small $ m_3 $  is achieved for the $ \mathcal{CP} $ phase parameters $\delta_\mathcal{CP} = 0$ and $\alpha_{21} = \pi$. It takes the following value,
\begin{equation}
|m^*_{ee}m_{\mu e}|\simeq\left|\Delta m^2_{31}\right| c^3_{13}\left(\frac{1}{2}c_{23}\sin 4\theta_{12} +s_{23}s_{13}\cos^22\theta_{12}\right)\simeq 6.1\cdot 10^{-4} \textrm{~eV}^2 ~,
\label{IO_max}
\end{equation}
while the minimal case occurs at the parameter set  described by the relation in Eq. (\ref{IO_sup}), for which we have $ |m^*_{ee}m_{\mu e}|=9.9 \cdot 10^{-6} $ eV$ ^2 $.

Using the result in Eq. (\ref{IO_max}), we determine the current and future expected lower bounds on the cubic coupling term $ \mu $,
\begin{equation}
\mu > 1.9 \cdot 10^{-7} \textrm{~GeV}\frac{m_{\Delta}}{800 \textrm{~GeV}}~,~~\textrm{and}~~
\mu > 1.9 \cdot 10^{-6} \textrm{~GeV}\frac{m_{\Delta}}{800 \textrm{~GeV}}~,
\end{equation}
respectively, with the corresponding bounds on the largest neutrino Yukawa coupling,
\begin{equation}
y< 0.0055 \frac{m_{\Delta}}{800 \textrm{~GeV}}~,~~\textrm{and}~~y< 0.00055\frac{m_{\Delta}}{800 \textrm{~GeV}}~. 
\label{y_con_mu3eIO}
\end{equation}

These maximal results are both significantly larger than the current best limit for the $ \mu\rightarrow e\gamma $ constraints found in Eq. (\ref{con_mueg3}), and both are stronger than the corresponding limits in the NO case. However for the minimal case in Eq. (\ref{IO_sup}) and for small $ m_3 $, the $\mu\rightarrow e\gamma $ constraints are dominant. A comparison of the current and future expected maximal reach on the triplet Higgs mass parameter is depicted in Figure \ref{IO_mu_3e}, where the above-described behaviour with changes in the $ \mathcal{CP} $ parameter sets can be seen.


\subsection{Current Constraints and Future Sensitivities for $\mu \rightarrow e$ Conversion in Nuclei Processes}
The LFV process of $\mu \rightarrow e$ Conversion in a nucleus, $ \mathcal{N} $, provides an important probe of the leptonic sector. In the context of the Type II Seesaw mechanism, this process is induced by the exchange of the doubly-charged component of the triplet Higgs and a loop contribution. The conversion rate is proportional to $\textrm{CR}(\mu\mathcal{N}\rightarrow e \mathcal{N})\propto
|C^{(II)}_{\mu e}|^2$, with the $ C^{(II)}_{\mu e} $  parameter taking the following form,
\begin{equation}
C^{(II)}_{\mu e}\equiv \frac{1}{4v^2_{\Delta}}\left[\frac{29}{24}\left(m^{\dagger}m\right)_{e\mu }
+\sum_{l=e,\mu,\tau} m^{\dagger}_{e l}f(r,s_l) m_{l\mu}
\right ]~,
\label{Cmue_eq}
\end{equation} 
where the loop contribution is described by the loop function $f(r,s_l)$, given by \cite{Raidal:1997hq},
\begin{equation}
f(r,s_l) = \frac{4s_l}{r}+\log(s_l)+\left(1-\frac{2s_l}{r}\right)\sqrt{1+\frac{4s_l}{r}}\log\frac{\sqrt{r+4s_l}+\sqrt{r}}{\sqrt{r+4s_l}-\sqrt{r}}~,
\end{equation}
in which $ r=m_\mu^2/m_\Delta^2 $ and $ s_l=m_l^2/m_\Delta^2 $ for our scenario. 

In contrast to the two LFV processes considered above, the  quantity constrained here ($ C^{(II)}_{\mu e} $) has an additional dependence on the triplet Higgs mass parameter through the loop function $ f(r,~s_l) $. In the limit of small $ m_l $ the loop function can be approximated by  $ f(r,s_l)\simeq \log(r) = \log(m_\mu^2/m_{\Delta}^2) $. Importantly, this loop function takes negative values for the range of $ m_\Delta $ parameters we consider, and thus leads to interesting behaviour as we vary $ m_\Delta $. Cancellations between the first and second term in Eq. (\ref{Cmue_eq}) can occur, suppressing the conversion rate to zero. Such behaviour will be illustrated in the next Section.

Currently, the best constraints on the conversion rate are provided by the SINDRUM experiment, which utilised Ti nuclei \cite{SINDRUMII:1993gxf}. This constraint translates to the following limit, 
\begin{equation}
|C^{(II)}_{\mu e}| < 7.94 \cdot 10^{-3}~
\left(\frac{m_{\Delta}}{800\textrm{~GeV}}\right)^{2}~,
\label{Climit1} 
\end{equation}
with a correspondingly a limit on the cubic coupling $ \mu $,
\begin{equation}
\mu > 2.4 \cdot 10^{-7} \textrm{~GeV}\frac{\sqrt{4v^2_{\Delta}|C^{(II)}_{\mu e}|}}{1 \textrm{~eV}} \frac{m_{\Delta}}{800 \textrm{~GeV}}~,
\end{equation}

The upcoming COMET experiment \cite{COMET:2018wbw,Moritsu:2022lem} is aiming to improve upon this limit to the following sensitivity,
\begin{equation}
4v^2_{\Delta}|C^{(II)}_{\mu e}| < 3.7 \cdot 10^{-6}\left(\frac{m_{\Delta}}{800\textrm{~GeV}}\right)^{2}~,
\label{Climit2}
\end{equation}
with a correspondingly a limit on the cubic coupling $ \mu $,
\begin{equation}
\mu > 1.1 \cdot 10^{-5} \textrm{~GeV}\frac{\sqrt{4v^2_{\Delta}|C^{(II)}_{\mu e}|}}{1 \textrm{~eV}} \frac{m_{\Delta}}{800 \textrm{~GeV}}~,
\end{equation}
which can be reinterpreted into a bound on the largest neutrino Yukawa coupling,
\begin{equation}
y< 9.6 \cdot 10^{-5} \frac{1 \textrm{~eV}}{\sqrt{4v^2_{\Delta}|C^{(II)}_{\mu e}|}} \frac{m_{\Delta}}{800 \textrm{~GeV}}~. 
\label{y_con_muecon}
\end{equation}
These projected bounds represent an approximate $ 46\times $ enhancement in the experimental sensitivity to the triplet Higgs cubic coupling and mass parameter.


\subsubsection*{Sensitivities for NO: $\mu \rightarrow e$ Conversion in Nuclei}

The  $\mu \rightarrow e$ conversion in nuclei process is significantly less sensitive to the choice of neutrino $\mathcal{CP}$ phases compared to the $\mu \rightarrow 3e$ decay, particularly in the NO case. This feature is important, as together with tests of the  $\mu \rightarrow 3e$ process, it becomes possible to pinpoint the  $\mathcal{CP}$ phases alongside the properties of the triplet Higgs parameters.

\begin{figure}[b]
\centering
\includegraphics[width=0.6\textwidth]{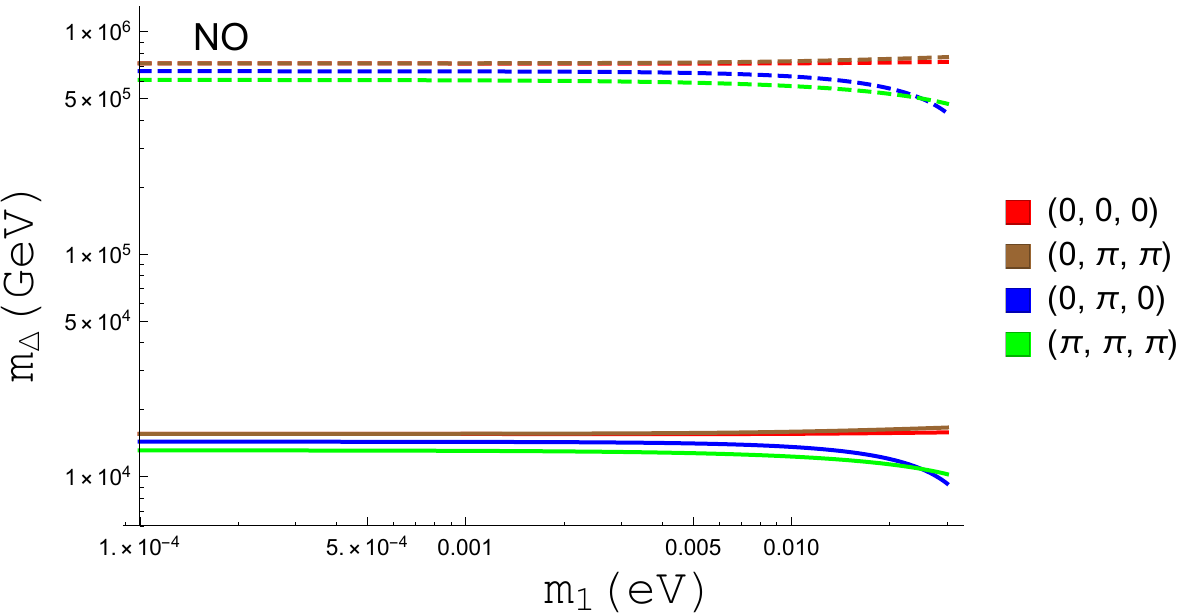}
\caption{Comparison plot of the maximum $ m_\Delta $ probed by  $ \mu $ to $ e $ conversion in Ti nuclei tests at the  SINDRUM experiment \cite{SINDRUMII:1993gxf} (Dashed) and at the upcoming COMET experiment \cite{Moritsu:2022lem} (Solid) for different sets of neutrino $\mathcal{CP}$ phases ($ \delta_{\mathcal{CP}} $, $ \alpha_{21} $, $ \alpha_{31} $) and varying lightest neutrino mass, in the NO scenario. All other parameters are given by the best fit parameters in Table \ref{nufit_params}.}
\label{NO_mu_e}
\end{figure}

As with the previously considered decay processes, we can determine the maximum and minimum values of the relevant parameter, namely $ |C^{(II)}_{\mu e}| $ . In the case of small $ m_1 $, this parameter is maximised for the $\mathcal{CP}$ phase parameter sets ($ 0 $, $ 0 $, $ 0 $) and ($ 0 $, $\pi$, $\pi$). Varying $ m_\Delta $ between $ 800 $ and $ 10^6 $ GeV, the maximum value is given by  $ 3.6\cdot 10^{-3}~\textrm{eV}^2~<|4v^2_{\Delta}C^{(II)}_{\mu e}|<7.3\cdot 10^{-3} $ eV$ ^2 $ respectively for the best fit values given in Table \ref{nufit_params}.

As discussed above, the fact that the loop function is negative allows for the possibility of cancellation and subsequent suppression of the conversion rate. An example of when this occurs is for the $\mathcal{CP}$ phase parameter set ($ 0 $, $\pi$, $0$) with triplet Higgs mass $ m_\Delta = 1000 $ GeV and lightest neutrino mass $ m_1\simeq 0.0256 $ eV. In the next section, this behaviour will be depicted, showing the movement of this zero with respect to $ m_\Delta $ and $ m_1 $.


\subsubsection*{Sensitivities for IO: $\mu \rightarrow e$ Conversion in Nuclei}

\begin{figure}[b]
\centering
\includegraphics[width=0.6\textwidth]{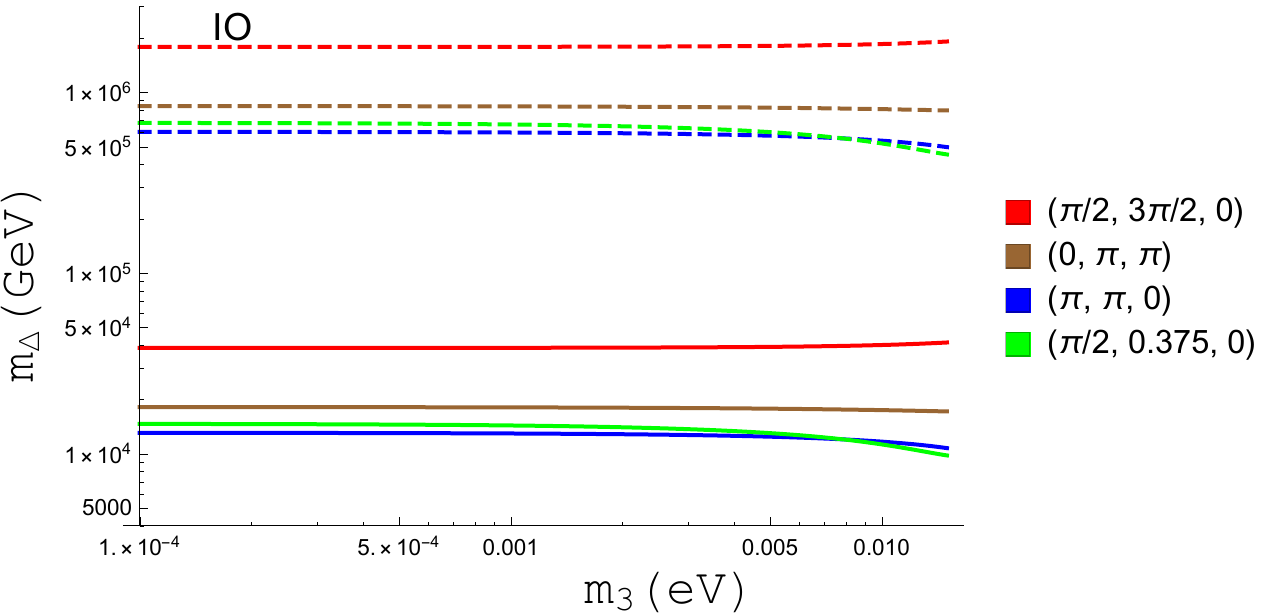}
\caption{Comparison plot of the maximum $ m_\Delta $ probed by  $ \mu $ to $ e $ conversion in Ti nuclei tests at the  SINDRUM experiment \cite{SINDRUMII:1993gxf} (Dashed) and at the upcoming COMET experiment \cite{Moritsu:2022lem} (Solid)  for different sets of neutrino $\mathcal{CP}$ phases ($ \delta_{\mathcal{CP}} $, $ \alpha_{21} $, $ \alpha_{31} $) and varying lightest neutrino mass, in the IO scenario. All other parameters are given by the best fit parameters in Table \ref{nufit_params}.}
\label{IO_mu_e}
\end{figure}

In comparison to the NO case, the IO exhibits greater variation with changes in the neutrino $\mathcal{CP}$ phases, but it remains less significant than the $\mu \rightarrow 3e$ decay process. Once again, through this feature, the combination of these two decay processes will allow the determination of the $\mathcal{CP}$ phases if there exists a triplet Higgs within the sensitivity range of these experiments. 

For the $\mathcal{CP}$ phase parameter set ($ \pi/2 $, $ 3\pi/2 $, $ 0 $), in the small $ m_3 $ limit, we obtain the maximum value of the $4v^2_{\Delta}|C^{(II)}_{\mu e}|$ parameter. Varying $ m_\Delta $ between $ 800 $ and $ 2\cdot 10^6 $ GeV, the maximum value is given by  $ 2.3\cdot 10^{-2}~\textrm{eV}^2~<|4v^2_{\Delta}C^{(II)}_{\mu e}|<4.6\cdot 10^{-2} $ eV$ ^2 $ respectively for the best fit values given in Table \ref{nufit_params}. These are almost an order of magnitude greater than for the NO scenario, meaning that once again the experimental tests of the LFV decay process are more sensitive to IO rather than NO. 

The minimum value of $4v^2_{\Delta}|C^{(II)}_{\mu e}|$ for small $ m_3 $ masses is found when taking the $\mathcal{CP}$ phase parameter set ($ \pi $, $ \pi $, $ 0 $). Varying $ m_\Delta $ between $ 800 $ and $ 2\cdot 10^6 $ GeV, the maximum value is given by  $ 1.3\cdot 10^{-3}~\textrm{eV}^2~<|4v^2_{\Delta}C^{(II)}_{\mu e}|<6.0\cdot 10^{-3} $ eV$ ^2 $, representing approximately an order of magnitude suppression. 
The $ m_\Delta $ dependent cancellation effect in the IO scenario will be seen in the next section for the $\mathcal{CP}$ phase parameter set ($ \pi $, $ \pi$, $ 0 $).


\subsection{Comparison of the Current Constraints and Future Sensitivities}
\label{con_sum}

\begin{figure}[b]
\begin{subfigure}
\centering
\includegraphics[width=0.4925\textwidth]{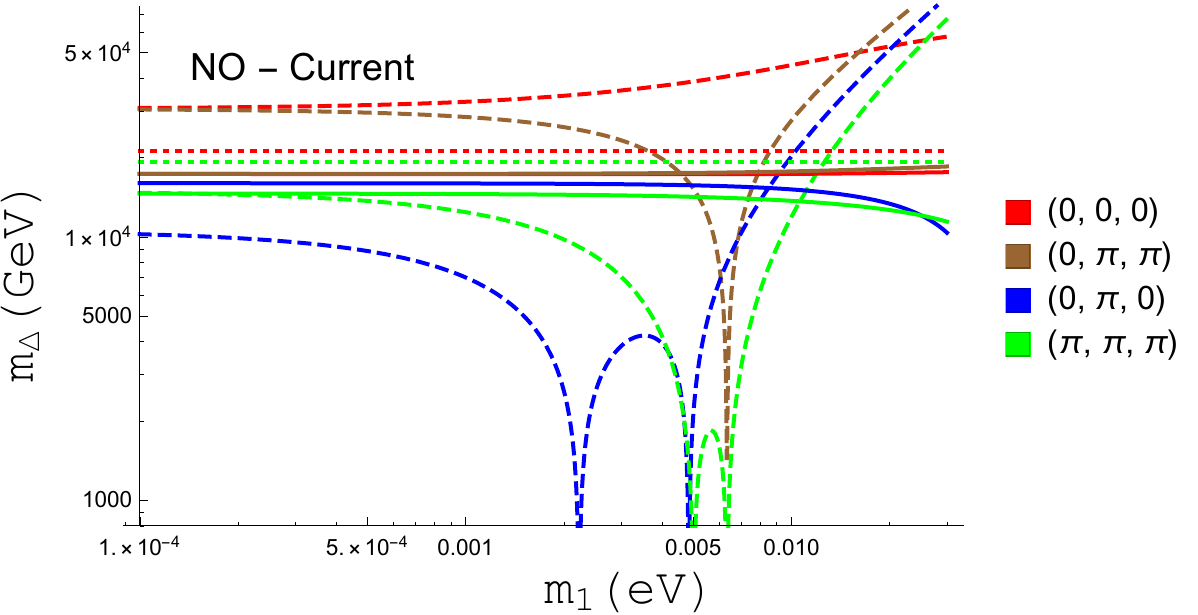}
\end{subfigure}
\hfill 
\begin{subfigure}
\centering
\includegraphics[width=0.4925\textwidth]{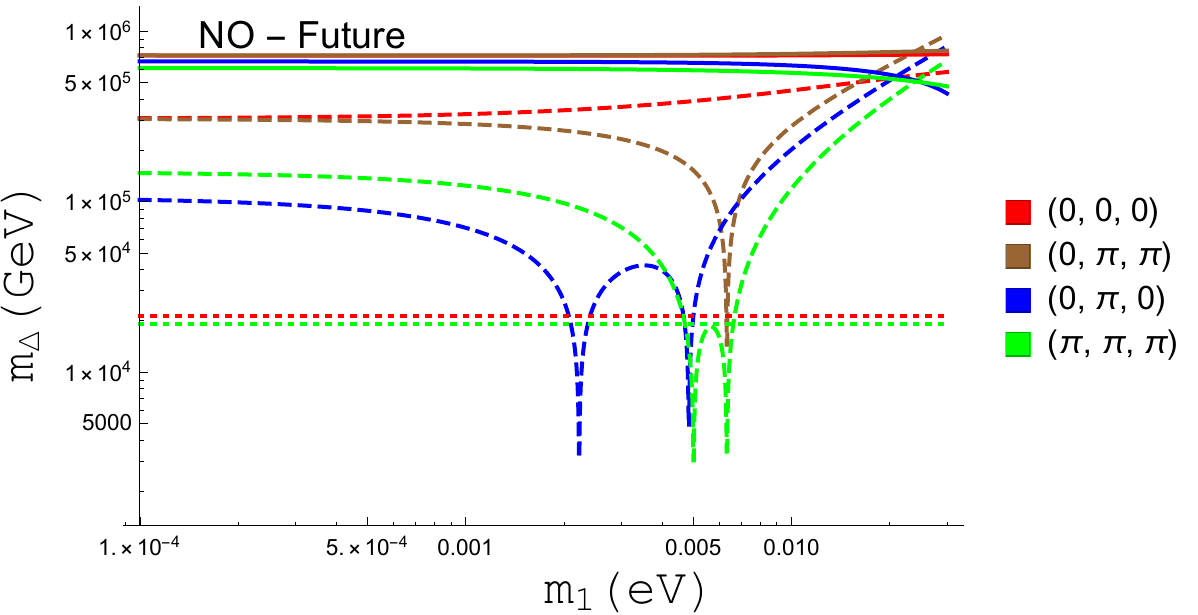}
\end{subfigure}
\begin{subfigure}
\centering
\includegraphics[width=0.4925\textwidth]{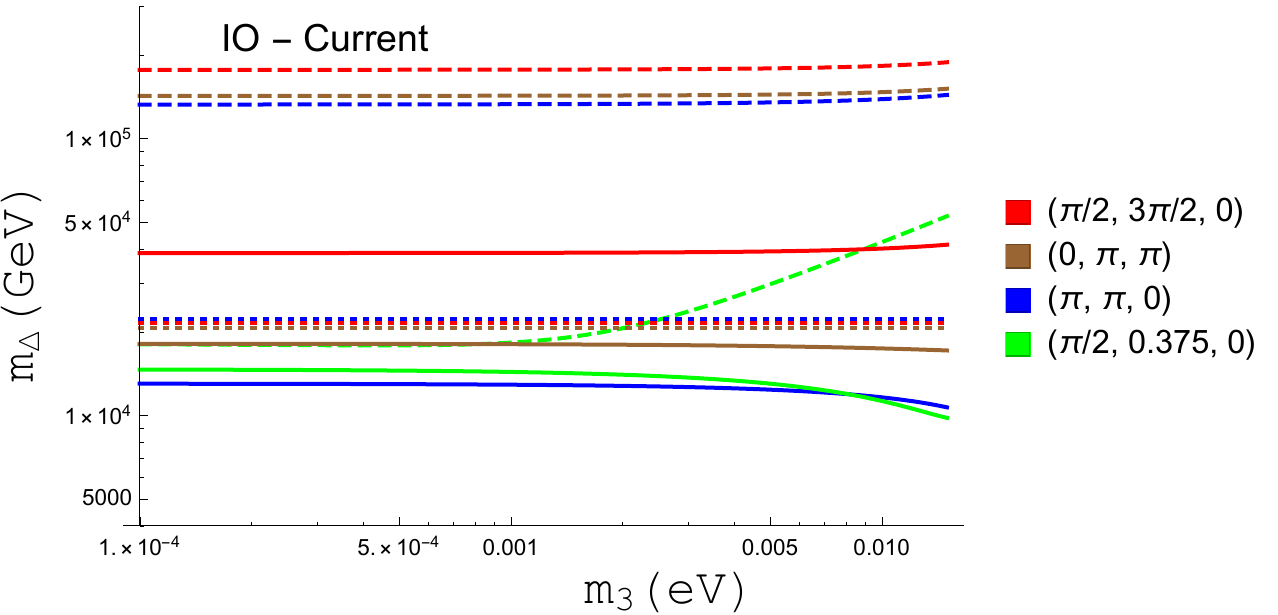}
\end{subfigure}
\hfill 
\begin{subfigure}
\centering
\includegraphics[width=0.4925\textwidth]{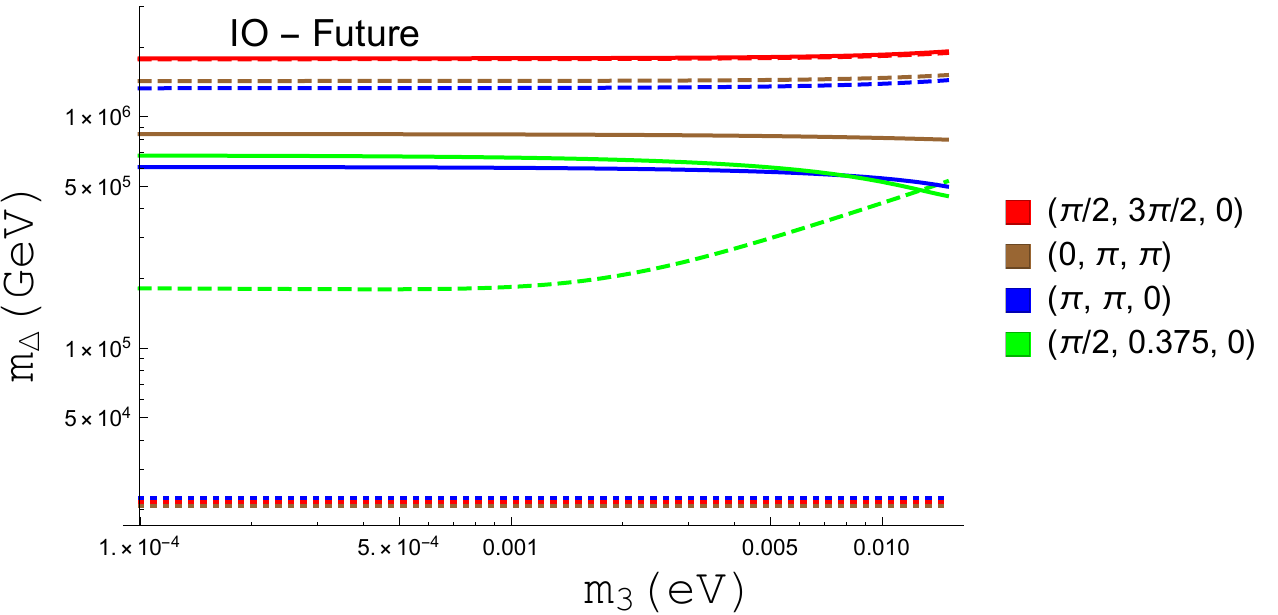}
\end{subfigure}
\caption{Comparison plot of the maximum expected $ m_\Delta $ probed by the $ \mu\rightarrow e\gamma $ (Dotted), $ \mu\rightarrow 3e $ (Dashed), and $ \mu $ to $ e $ conversion in Ti nuclei  (Solid) processes for different sets of neutrino $\mathcal{CP}$ phases ($ \delta_{\mathcal{CP}} $, $ \alpha_{21} $, $ \alpha_{31} $) and varying lightest neutrino mass. The current (Left) and future expected sensitivity (Right) are depicted in both the NO (Top) and IO (Bottom) scenarios. All other parameters are given by the best fit parameters in Table \ref{nufit_params}.}
\label{NOIO_com}
\end{figure}

Here we briefly summarise the results of this Section to illustrate the importance of each LFV decay process for testing Type II Seesaw Leptogenesis. In Figure \ref{NOIO_com}, a comparison is depicted of the sensitivities for each of the LFV processes discussed above, including both current and upcoming constraints. In the case of NO, there is a dependence on the $\mathcal{CP}$ phase parameter set for determining whether the $ \mu\rightarrow e\gamma $ decay measurements at the MEG experiment or the $ \mu\rightarrow 3e $ constraints from the SINDRUM experiment provide the strongest limits. Once the upcoming experiments reach their desired sensitivity, the $ \mu $ to $ e $ conversion in Ti nuclei process at the COMET experiment will be the most sensitive test of the Triplet Higgs properties, except for large $ m_1 $. 

For the IO scenario, the  MEG results for the $ \mu\rightarrow e\gamma $ decay process are already exceeded by the  $ \mu\rightarrow 3e $ decay process measurements undertaken at the SINDRUM experiment - except in the minimal case. Interestingly, the Mu3e experiments search for the $ \mu\rightarrow 3e $ decay process will provide the strongest results for much of the $ \mathcal{CP} $ phase parameter space, exceeding the expected reach of the measurements of the $ \mu $ to $ e $ conversion in Ti nuclei process by the COMET experiment. Both of these experiments provide significant improvements in sensitivity to the triplet Higgs properties over the MEG results for the $ \mu\rightarrow e\gamma $ decay process. 

Each of the LFV experiments have different sensitivities for each of the $\mathcal{CP}$ phase sets and for varied lightest neutrino mass. These differences will provide the means to pinpoint the  $\mathcal{CP}$ phases and the properties of the triplet Higgs parameters. This would not be possible through testing a single decay process.


\section{Expected Experimental Reach in the Allowed Parameter Space}
\label{ex_param}

Now that we have surveyed the current and expected future limits on the three LFV decay processes, we can determine how much of the allowed parameter space for successful Type II Seesaw Leptogenesis can be probed. The allowed parameter space can be depicted in the $ \mu $-$ m_\Delta $ parameter space, of which the main constraints for the mass range $ m_\Delta<2 \cdot 10^6 $ GeV are currently  the prevention of lepton asymmetry washout discussed in Section \ref{TIILeptp}, the requirement of perturbative Yukawa couplings up to the Planck scale, the LHC limit of $m_\Delta>800$ GeV, and the current LFV constraints. In each of the results figures presented below, these constraints are depicted. The Light Grey, Black, and Grey regions correspond to the lepton asymmetry washout, non-perturbative neutrino Yukawa coupling, and current LFV limits respectively. The region subtended by these constraints contains parameter values that lead to successful Leptogenesis, neutrino mass generation, and inflation. 

An example set of parameters for the Type II Seesaw Leptogenesis scenario that are consistent with this allowed region are non-minimal couplings of  $\xi_H=\xi_\Delta=300$, an initial phase of $\theta_{0}=0.1$, and couplings $\lambda_5=4\cdot 10^{-12},~ \lambda_H=0.1,~  \lambda_{H\Delta}=-0.001$, and $~ \lambda_{\Delta}=4.5 \cdot 10^{-5}$; see Ref. \cite{Barrie:2022cub} for more details regarding this choice of parameter set. It is important to note that this parameter set has not included the running of the couplings to the Planck scale, which is sensitive to the triplet Higgs parameters probed by LFV experiments. The limits derived from the LFV decays place bounds on these couplings, constraining the parameter space consistent with vacuum stability and the inflationary observables - establishing the key connection between the low energy LFV experimental signatures and the high scale Type II Seesaw Leptogenesis mechanism. 

The current LFV limits depicted in each of the figures are derived from the most constraining of the current bounds on the three lepton violating processes described above. As shown in Figure \ref{NOIO_com}, for some sets of neutrino $ \mathcal{CP} $ phases and lightest neutrino masses, the $ \mu\rightarrow 3e $ processes can be strongly suppressed such that the $ \mu\rightarrow e\gamma $ constraints provide the best limits. This is particularly relevant when the $ \mu\rightarrow 3e $ branching ratio is suppressed to zero for some parameter choices, pushing it well below the current bounds on both $ \mu\rightarrow e\gamma $ and $ \mu $ to $ e $ conversion in Nuclei processes. Suppression of the  $ \mu $ to $ e $ conversion process tends to be less uniformly severe across the $ m_\Delta $ range, but the current limits are significantly weaker than for the $ \mu\rightarrow e\gamma $ decay process.

In each Figure, the Dotted lines subtending Red regions and Dashed lines subtending Green regions denote the projected experimental sensitivity of the future COMET ($ \mu $ to $ e $ conversion in Nuclei) and Mu3e ($ \mu\rightarrow 3e $) experiments \cite{Moritsu:2022lem,Perrevoort:2018ttp}, respectively. The projected sensitivity of a future 100 TeV collider is given by the orange region, which will be able to probe the mass of the triplet Higgs up to 4 TeV at the $ 5\sigma $ level \cite{Du:2018eaw}. Note, that the branching ratios of the doubly-charged component of the Triplet Higgs are dependent on the mixing angles in the leptonic sector which may provide a complementary test \cite{Garayoa:2007fw}. In all figures, it is interesting to note the overlap of the projected sensitivities of the LFV processes and 100 TeV collider searches. Illustrating the complementary nature of each of these experimental approaches and the rich phenomenological implications of this model.

It is important to note that the constraints on the $ \mu $ and $ m_\Delta $ parameter space presented in this Section can be easily translated into limits on the maximal neutrino Yukawa coupling, as discussed in the previous Section. Such constraints are important for understanding the running of the various scalar couplings, and ensuring vacuum stability up to the Planck scale due to their sensitivity to the neutrino Yukawa couplings and $ m_\Delta $ parameter. Additionally, the determination of the dominant component of the inflaton, and subsequently which dimension five lepton violating interaction leads to Leptogenesis, is related to the relative size of the scalar and non-minimal couplings at the Planck scale. 

\begin{figure}[b!]
\begin{subfigure}
\centering
\includegraphics[width=0.4925\columnwidth]{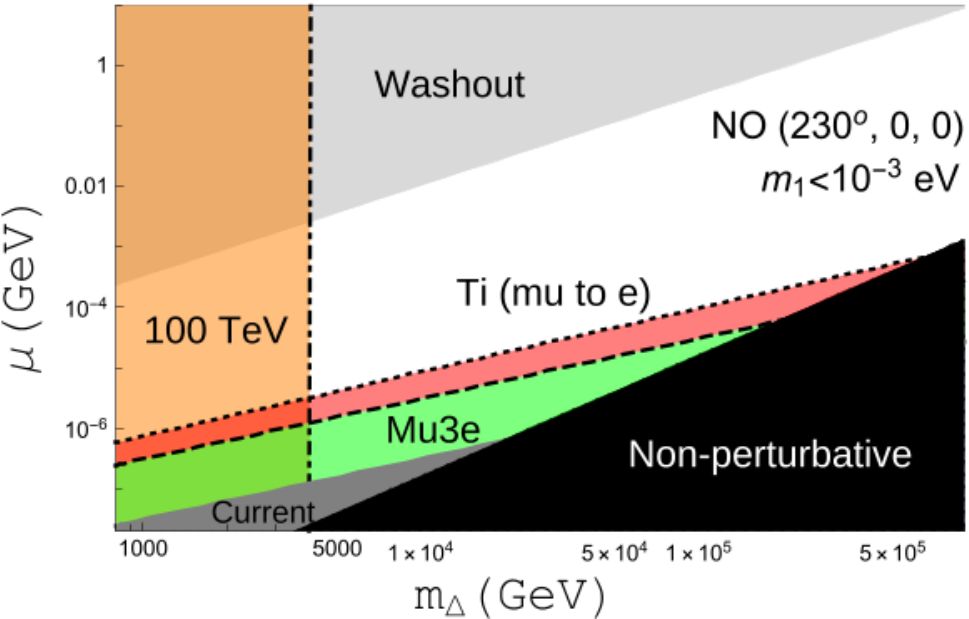}
\end{subfigure}
\hfill 
\begin{subfigure}
\centering
\includegraphics[width=0.4925\columnwidth]{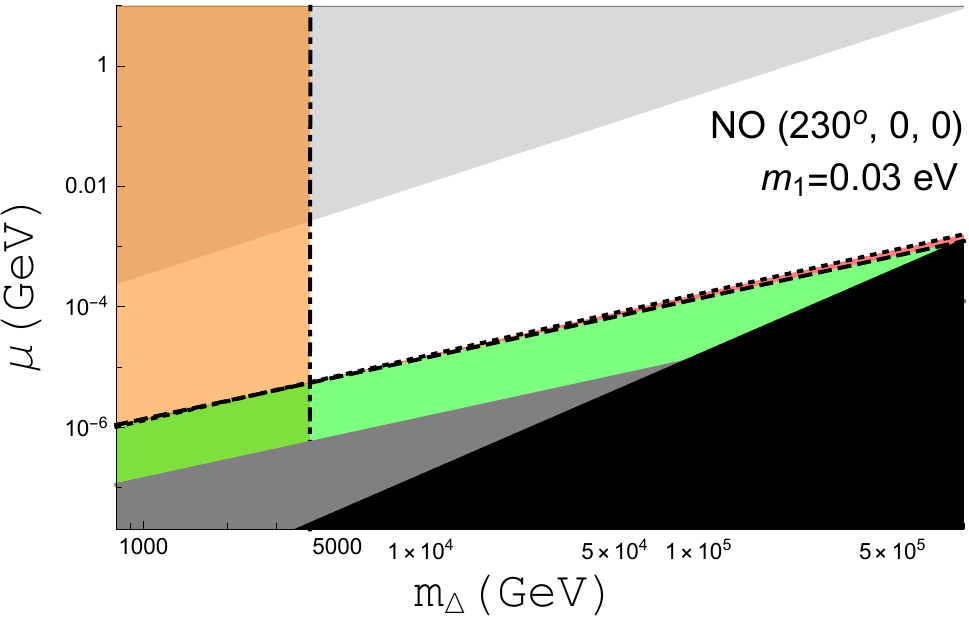}
\end{subfigure}
\begin{subfigure}
\centering
\includegraphics[width=0.4925\columnwidth]{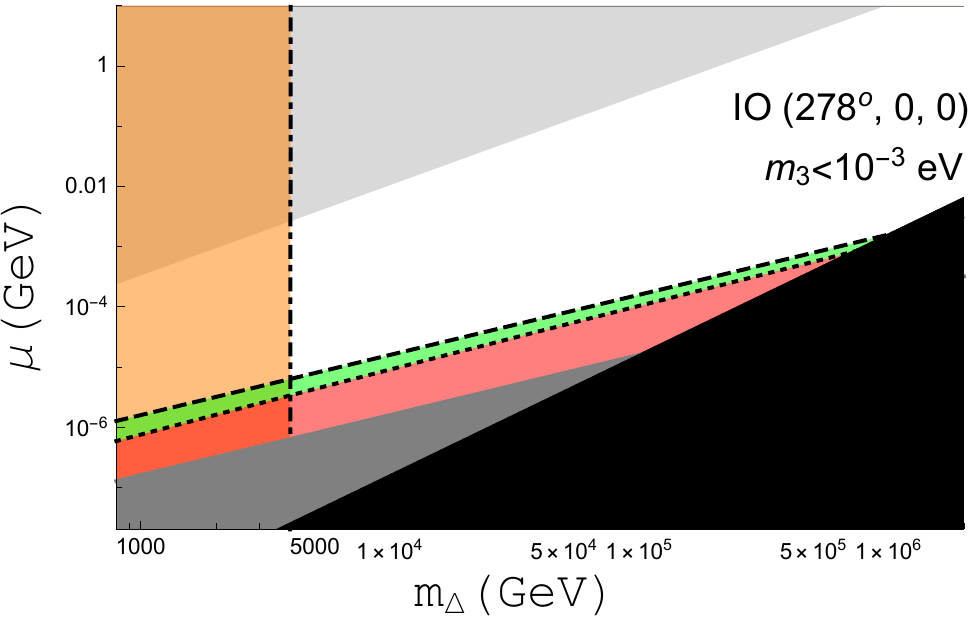}
\end{subfigure}
\hfill 
\begin{subfigure}
\centering
\includegraphics[width=0.4925\columnwidth]{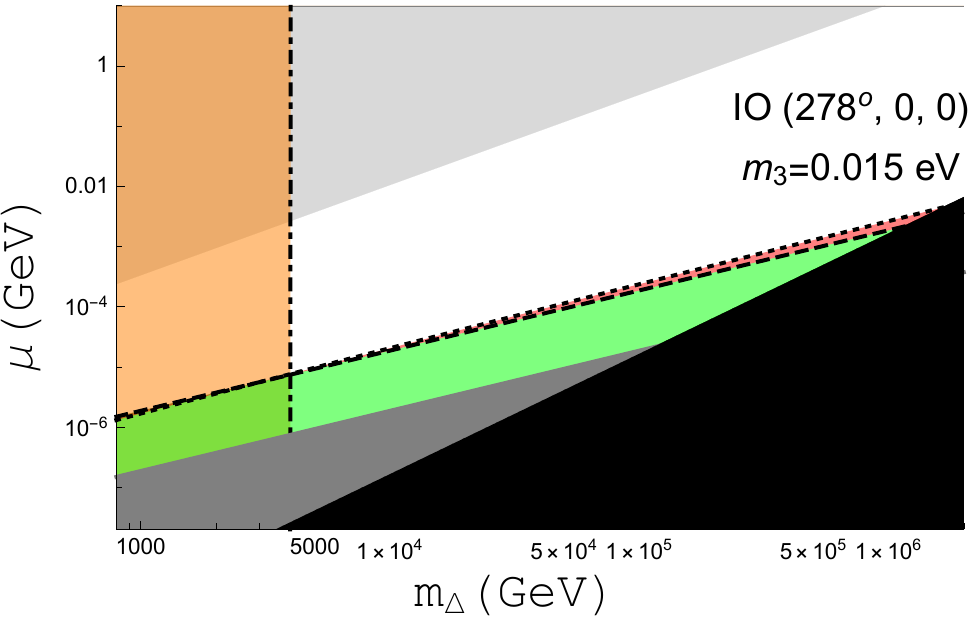}
\end{subfigure}
\caption{The reach of LFV searches is depicted for the current best fit parameters, given in Table \ref{nufit_params}, for different $ m_1 $ mass regimes and $\alpha_{21}=\alpha_{31}=0$. The upper bounds on the future sensitivities of the future Mu3e and $ \mu $ to $ e $ conversion experiments are denoted by the dashed line (subtending the green region) and dotted line (subtending the red region) respectively \cite{Perrevoort:2018ttp}.  The discovery potential of a future 100 TeV collider is depicted by the orange region, subtended by the black dot-dashed line \cite{Du:2018eaw}. Significant lepton asymmetry washout occurs for parameters within the Light Grey region, and non-perturbative neutrino Yukawa couplings (Black). The grey region denotes the current best limits either from $\mu\rightarrow 3e$ or $\mu\rightarrow e\gamma$ constraints, depending on the neutrino parameters \cite{SINDRUM:1987nra}. The region subtended by these three constraints is the parameter region where successful Leptogenesis, neutrino mass generation, and inflation can occur. The other neutrino parameters are given by the best fit parameters in Table \ref{nufit_params}. Both the NO (Top) and IO (Bottom) scenarios are included.}
\label{NOIO_bfp} 
\end{figure}


\subsection{Experimental Sensitivity in the NO scenario}

Firstly, we survey the allowed parameter regions and projected reach of upcoming experiments in the NO scenario, for different sets of neutrino $\mathcal{CP}$ phases and lightest neutrino masses. In Figure \ref{NOIO_bfp}, the best fit parameters for small and large $ m_1 $ are depicted, including the current $ \delta_{\mathcal{CP}} $ phase of $ 230^\circ $ with the Majorana phases set to $\alpha_{21}=\alpha_{31}=0$. 

\begin{figure}[b]
\begin{subfigure}
\centering
\includegraphics[width=0.4925\textwidth]{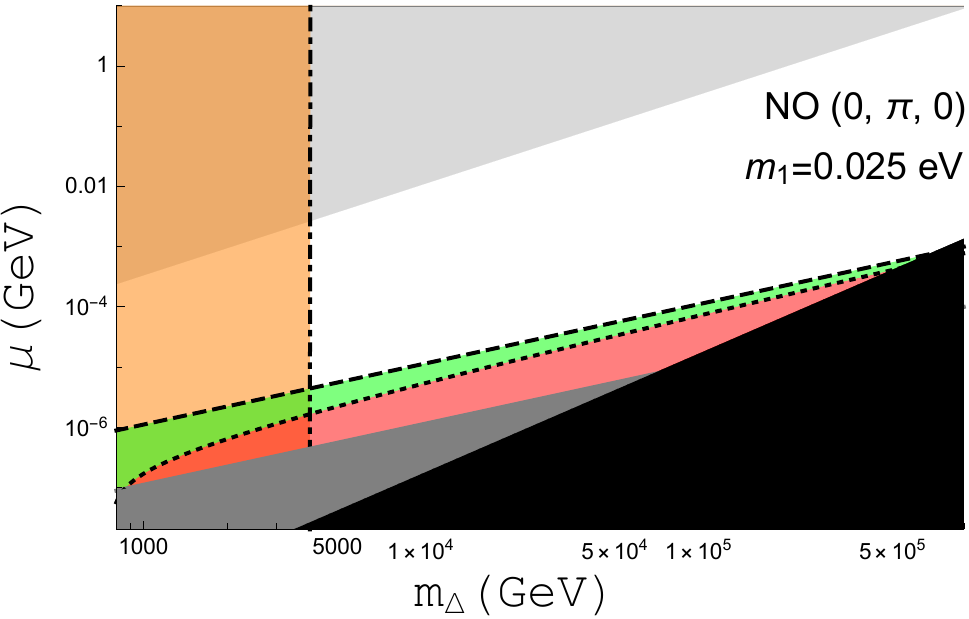}
\end{subfigure}
\hfill 
\begin{subfigure}
\centering
\includegraphics[width=0.4925\textwidth]{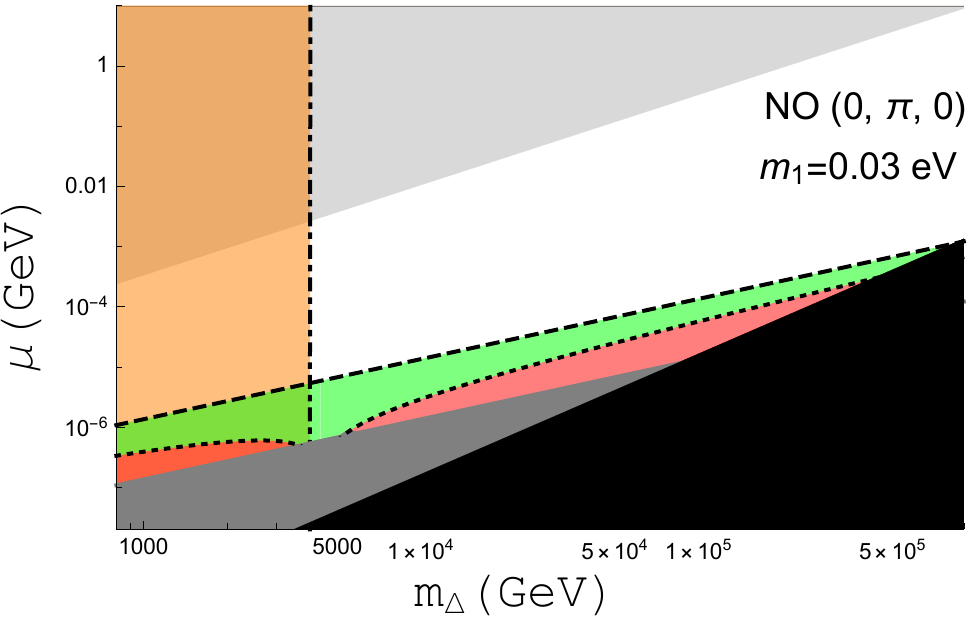}
\end{subfigure}
\begin{subfigure}
\centering
\includegraphics[width=0.4925\textwidth]{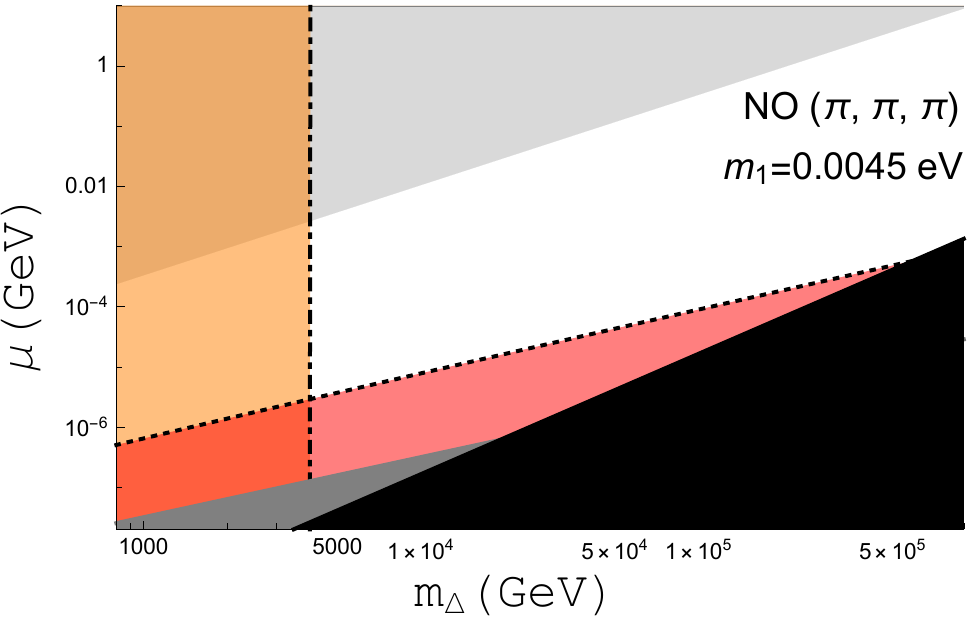}
\end{subfigure}
\hfill 
\begin{subfigure}
\centering
\includegraphics[width=0.4925\textwidth]{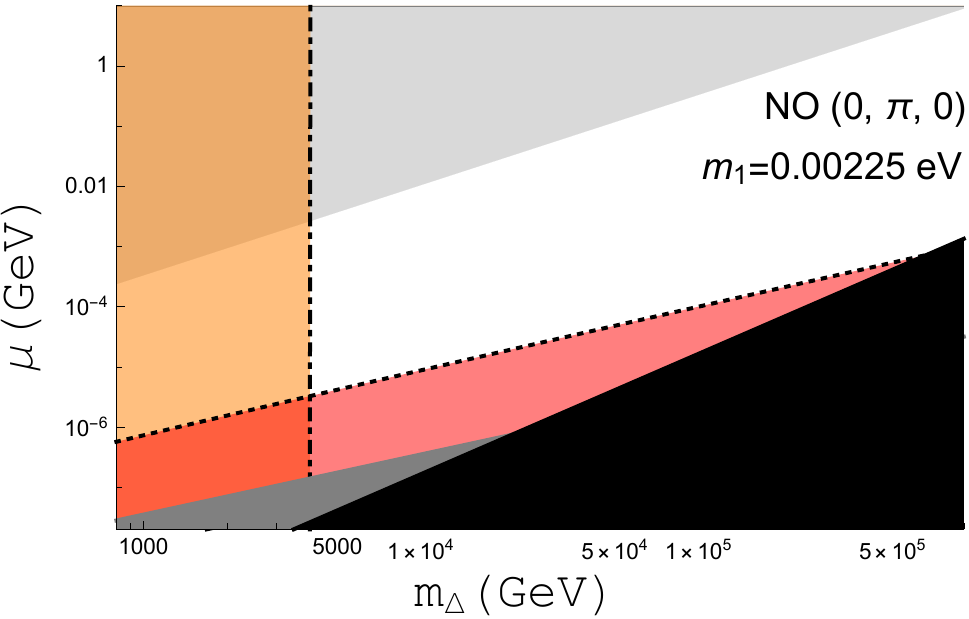}
\end{subfigure}
\caption{ The experimental sensitivities of LFV searches for the successful Type II Seesaw Leptogenesis parameter space is depicted for different $\mathcal{CP}$ phases parameter sets ($ \delta_{\mathcal{CP}} $, $ \alpha_{21} $, $ \alpha_{31} $) in the NO scenario with varying $ m_1 $ around special features.}
\label{NO_mass_var}
\end{figure}

In Figure \ref{NO_mass_var}, the suppression effects present in the $ \mu $ to $ e $ conversion in nuclei conversion rate and  the $ \mu\rightarrow 3e $ branching ratio are demonstrated. The top two figures show the $ m_\Delta $ dependent cancellation in Eq. (\ref{Cmue_eq}), and how it changes with choice of $ m_1 $. The bottom two figures depict two separate scenarios in which the $ \mu\rightarrow 3e $ branching ratio is suppressed such that no constraints are applied on the allowed parameter space by this process. This suppression is uniform across the $ m_\Delta $ range, in contrast to the cancellation seen for the $ \mu $ to $ e $ conversion in nuclei process.

\begin{figure}[t]
\begin{subfigure}
\centering
\includegraphics[width=0.4925\textwidth]{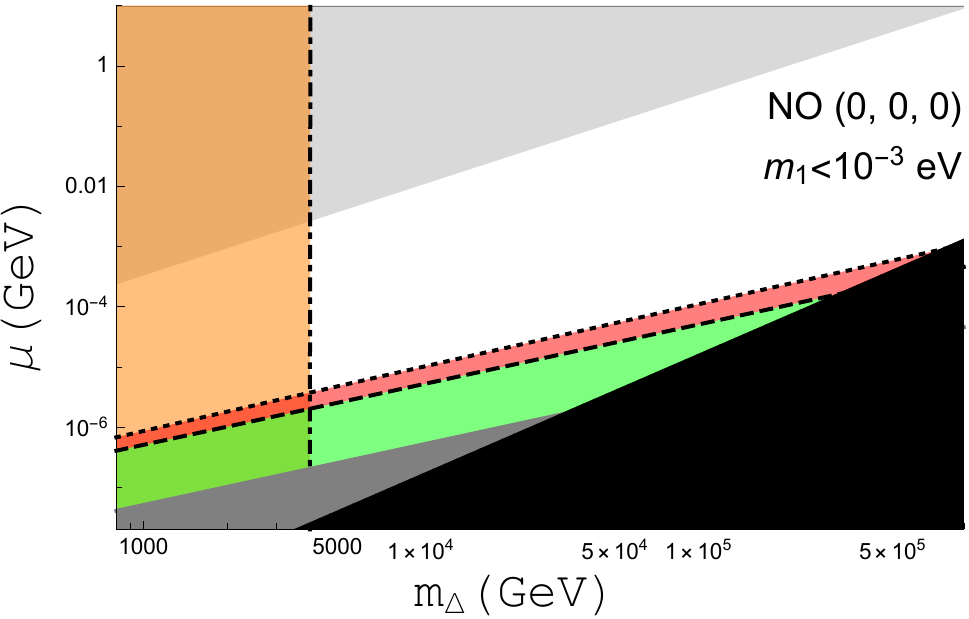}
\end{subfigure}
\hfill 
\begin{subfigure}
\centering
\includegraphics[width=0.4925\textwidth]{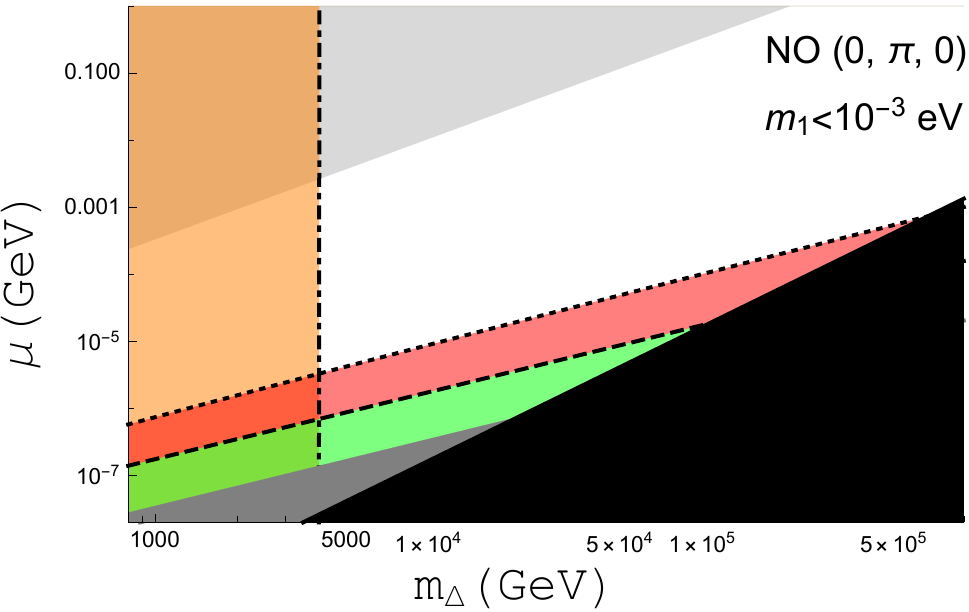}
\end{subfigure}
\begin{subfigure}
\centering
\includegraphics[width=0.4925\textwidth]{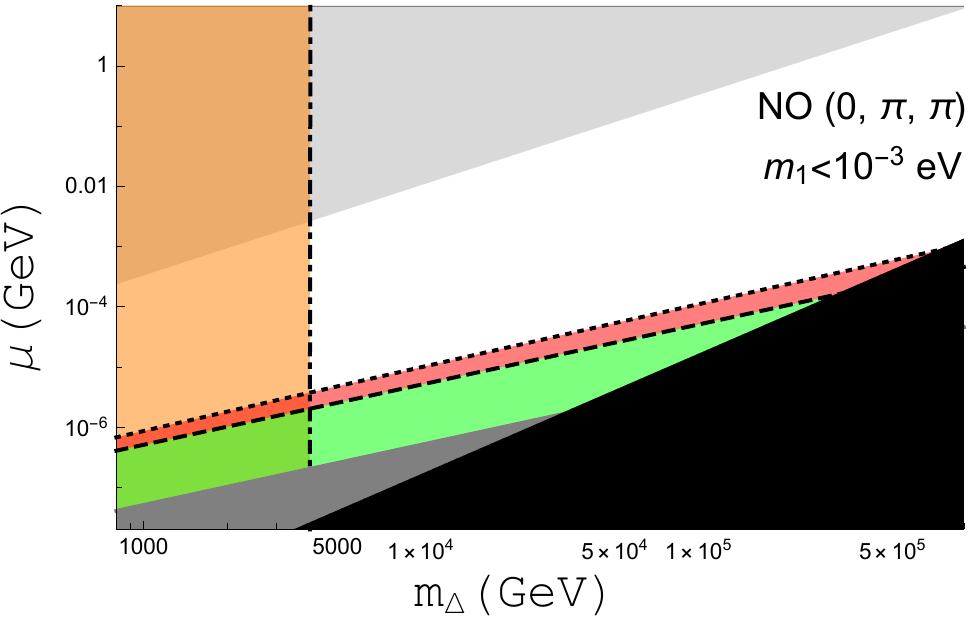}
\end{subfigure}
\hfill 
\begin{subfigure}
\centering
\includegraphics[width=0.4925\textwidth]{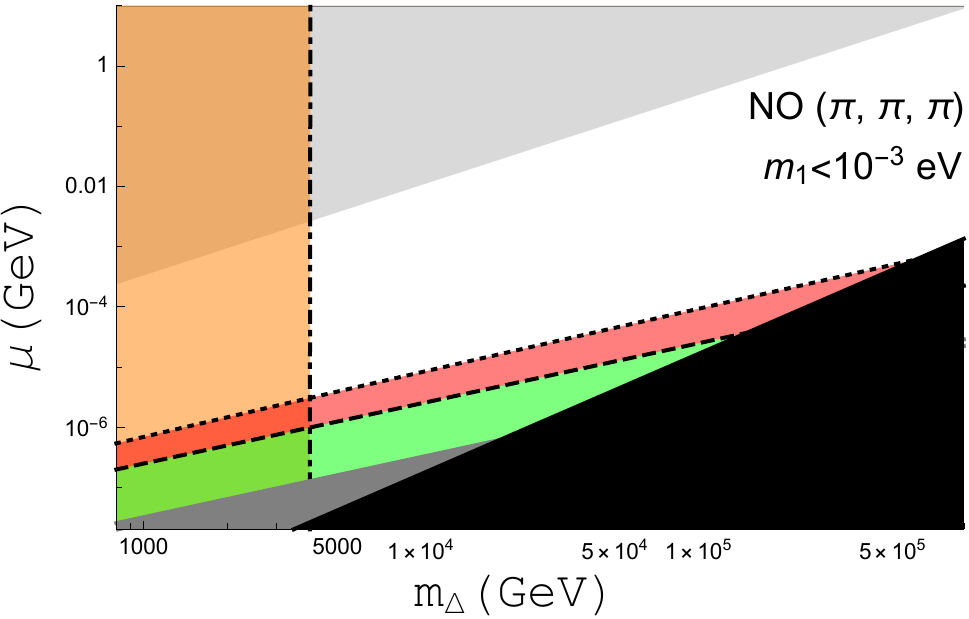}
\end{subfigure}
\caption{The experimental sensitivities of LFV searches for the successful Type II Seesaw Leptogenesis parameter space is depicted for different $\mathcal{CP}$ phases parameter sets ($ \delta_{\mathcal{CP}} $, $ \alpha_{21} $, $ \alpha_{31} $) in the NO scenario with small $ m_1 $.} 
\label{NO_angle_var}
\end{figure}

Figure \ref{NO_angle_var} illustrates the effects of different $\mathcal{CP}$ phase parameter sets for a small lightest neutrino mass. As expected from Section \ref{LFVs}, the $ \mu $ to $ e $ conversion in Ti nuclei experimental reach does not change significantly, while the $ \mu\rightarrow 3e $ limit varies. In each case, the $ \mu $ to $ e $ conversion in Ti nuclei at the upcoming COMET experiment provides the greatest sensitivity to the triplet Higgs parameters.


\subsection{Experimental Sensitivity in the IO scenario}

Now we consider the projected experimental sensitivity to the IO scenario for different sets of neutrino $\mathcal{CP}$ phases and the lightest neutrino mass. Beginning with the best fit parameters including the current $ \delta_{\mathcal{CP}} $ phase, which is depicted in Figure \ref{NOIO_bfp} for small and large $ m_3 $ mass parameter choices and $\alpha_{21}=\alpha_{31}=0$~.

\begin{figure}
\begin{subfigure}
\centering
\includegraphics[width=0.4925\textwidth]{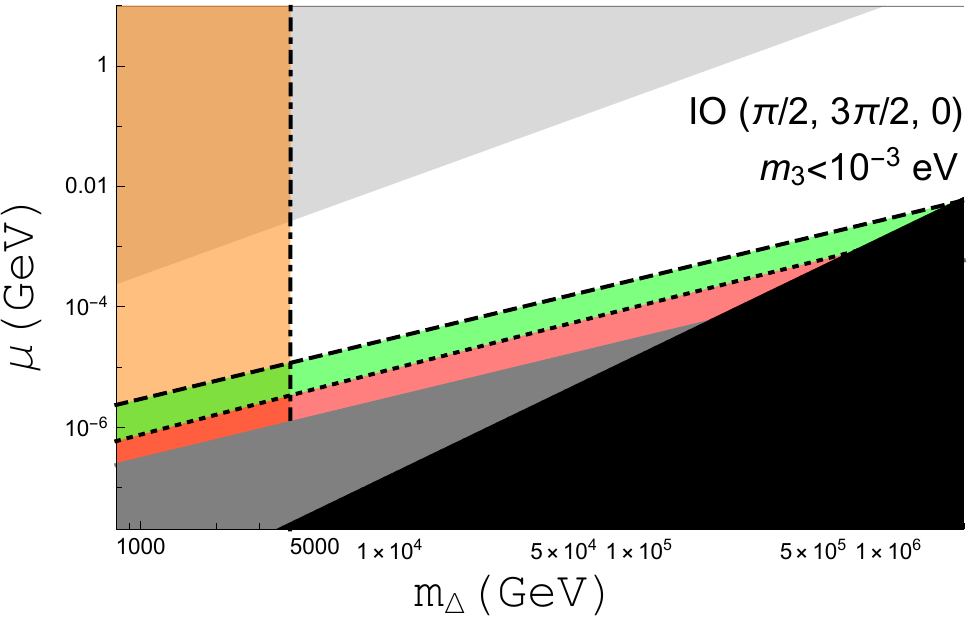}
\end{subfigure}
\hfill
\begin{subfigure}
\centering
\includegraphics[width=0.4925\textwidth]{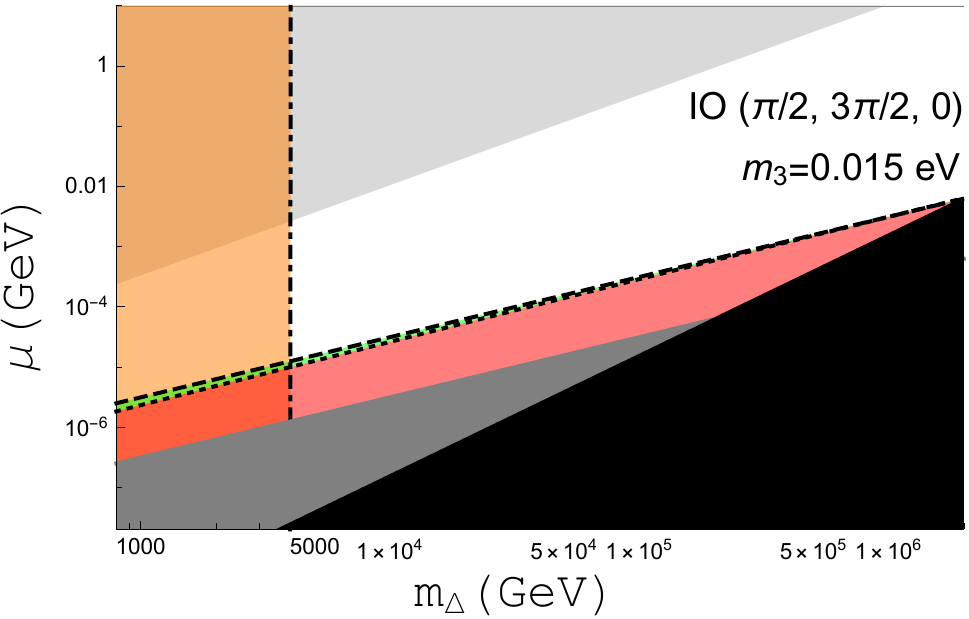}
\end{subfigure} 
\begin{subfigure}
\centering
\includegraphics[width=0.4925\textwidth]{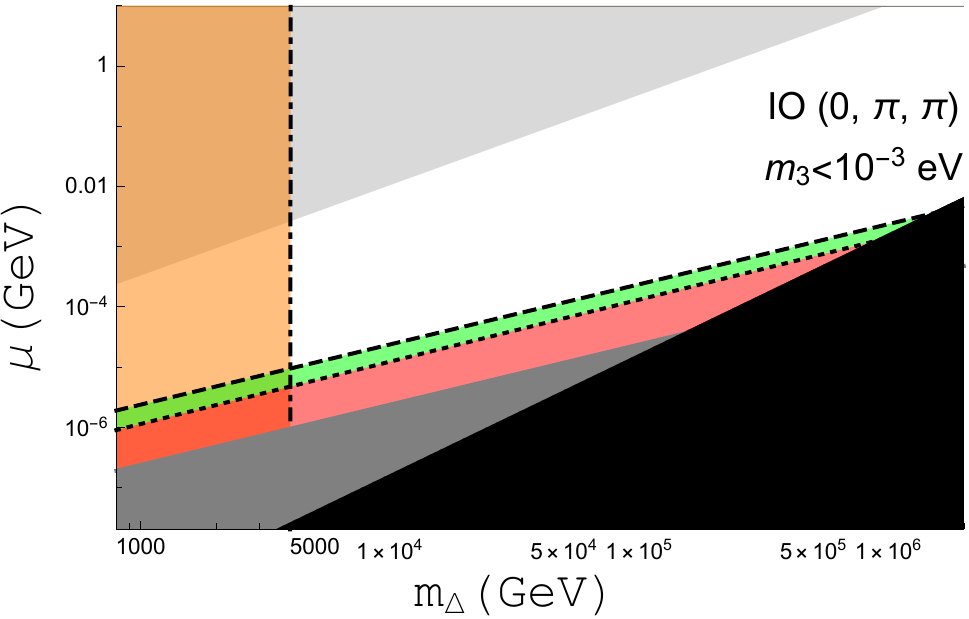}
\end{subfigure}
\hfill 
\begin{subfigure}
\centering
\includegraphics[width=0.4925\textwidth]{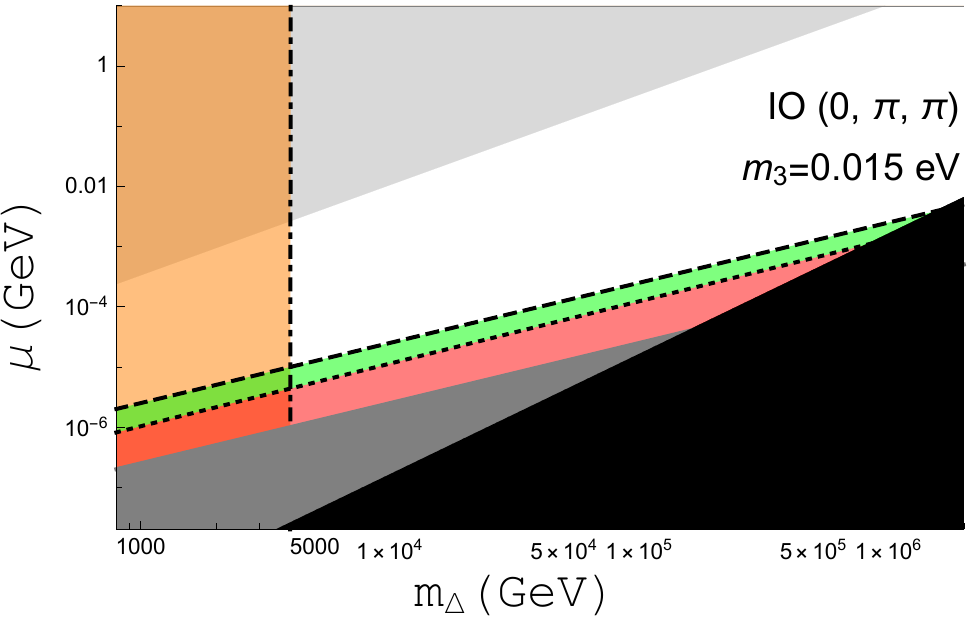}
\end{subfigure}
\begin{subfigure}
\centering
\includegraphics[width=0.4925\textwidth]{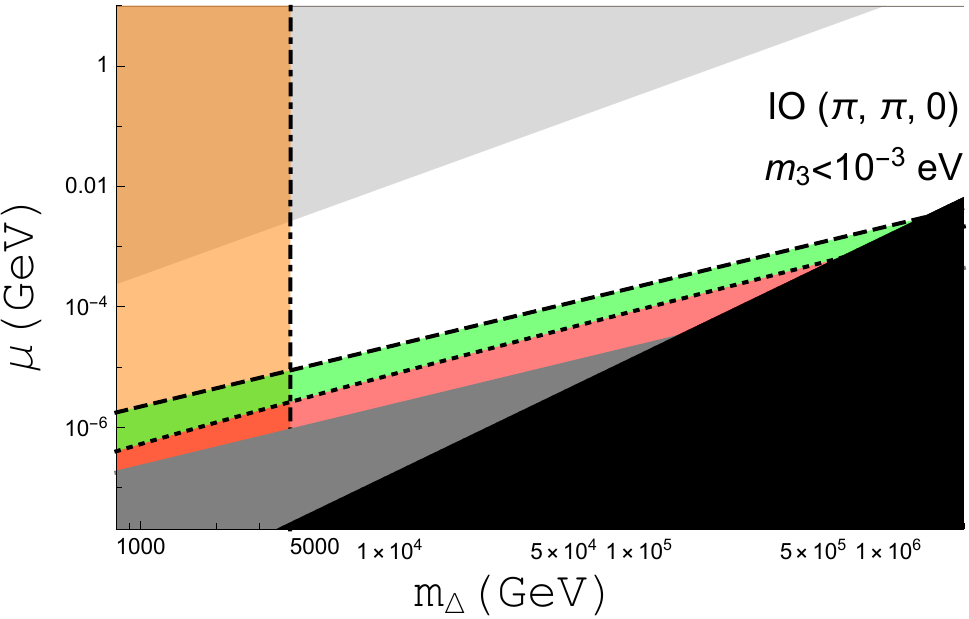}
\end{subfigure}
\hfill 
\begin{subfigure}
\centering
\includegraphics[width=0.4925\textwidth]{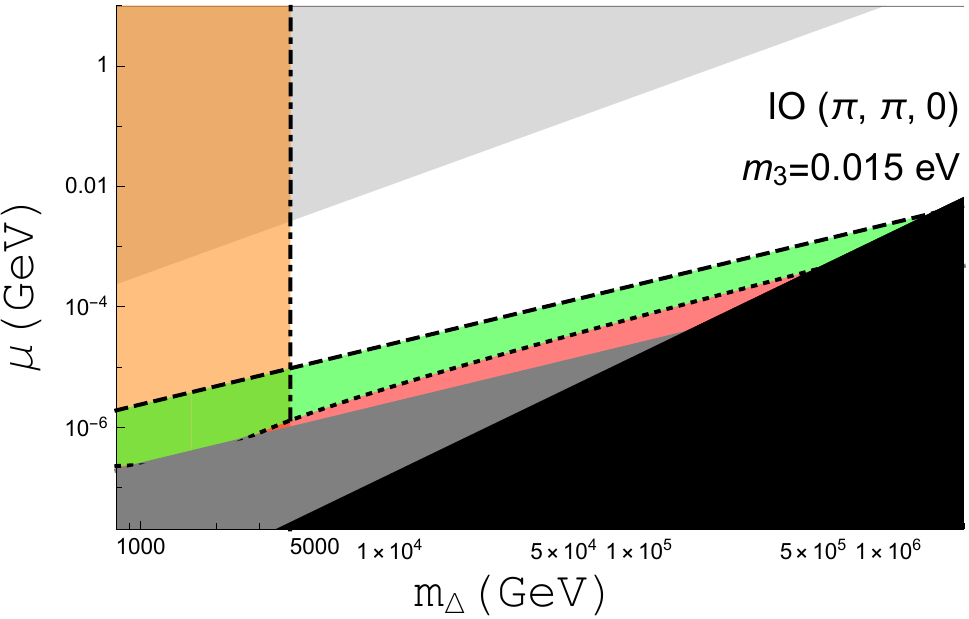}
\end{subfigure}
\begin{subfigure}
\centering
\includegraphics[width=0.4925\textwidth]{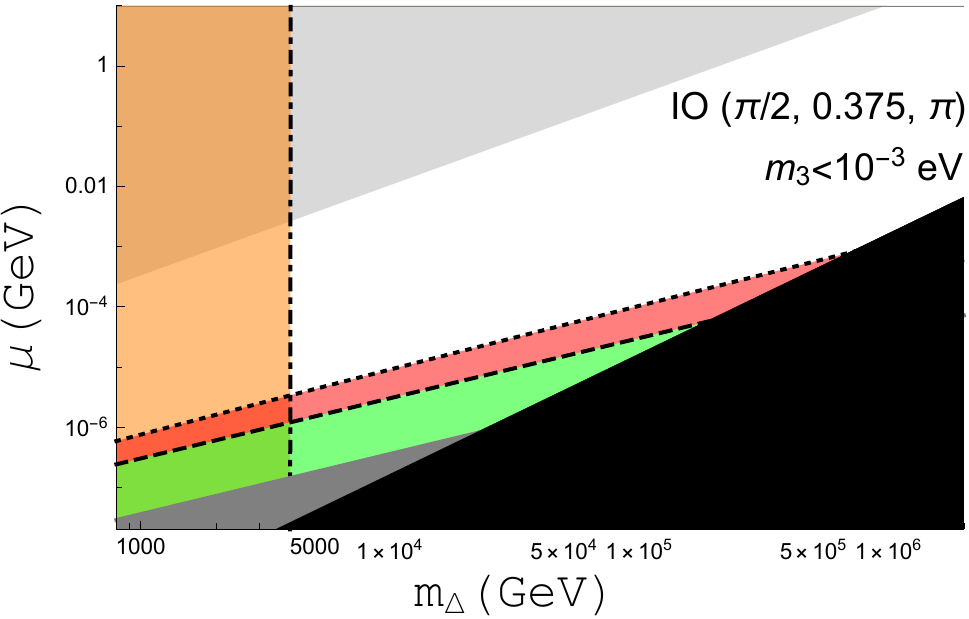}
\end{subfigure}
\hfill 
\begin{subfigure}
\centering
\includegraphics[width=0.4925\textwidth]{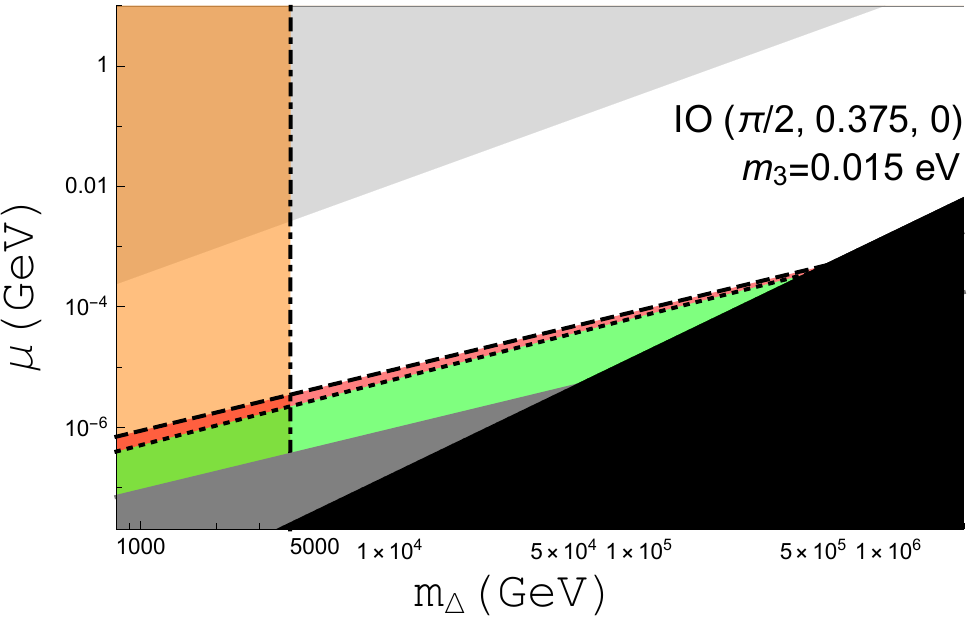}
\end{subfigure}
\caption{The experimental sensitivities of LFV searches for the successful Type II Seesaw Leptogenesis parameter space is depicted for different $\mathcal{CP}$ phases parameter sets ($ \delta_{\mathcal{CP}} $, $ \alpha_{21} $, $ \alpha_{31} $) in the IO scenario with varying $ m_1 $ inputs and $\mathcal{CP}$ parameter sets.}
\label{IO_mass_var}
\end{figure}

Figure \ref{IO_mass_var} exhibits the results for different $ \mathcal{CP} $ phase parameter sets for varied choices of the lightest neutrino mass. Significant variation in the projected constraints and dominant decay process is seen across each of these cases. Interestingly, in the ($ \pi $, $ \pi $, $ 0 $) parameter set there occurs a $ m_\Delta $ dependent cancellation in the conversion rate parameter in Eq. (\ref{Cmue_eq}), with subsequent suppression of the  $ \mu $ to $ e $ conversion rate. In contrast to the NO scenario, the projected limits from $\mu\rightarrow 3e$ decay processes tend to dominate over the  $ \mu $ to $ e $ conversion in Ti nuclei process in the selected $ \mathcal{CP} $ phase parameter sets. The exception to this is for the $ \mathcal{CP} $ phase parameter set that minimises the $\mu\rightarrow 3e$ branching ratio, namely ($\pi/2$, $ 0.375 $, 0) as found in Eq. (\ref{IO_sup}). Overall, the IO scenario generally provides a stronger test of the allowed parameter space for successful Type II Seesaw Leptogenesis.

\subsection{Neutrinoless Double Beta Decay Predictions}

To conclude this Section, we wish to also comment on the complementary test that is provided by neutrinoless double beta decay searches. Neutrinoless double beta decay is a process predicted in models that contain Majorana mass terms for the neutrinos, which may be testable in the near future. Theories that provide unique predictions of this process may be differentiated or ruled out by increased experimental precision in upcoming experiments. In the Type II Seesaw mechanism, the neutrino masses generated by the  triplet Higgs vacuum expectation value are of the Majorana type. Thus, this would be an expected experimental signal of our model, which is dependent upon the mixing properties of the neutrino sector.

\begin{figure}[h]
\begin{subfigure}
\centering
\includegraphics[width=0.4925\textwidth]{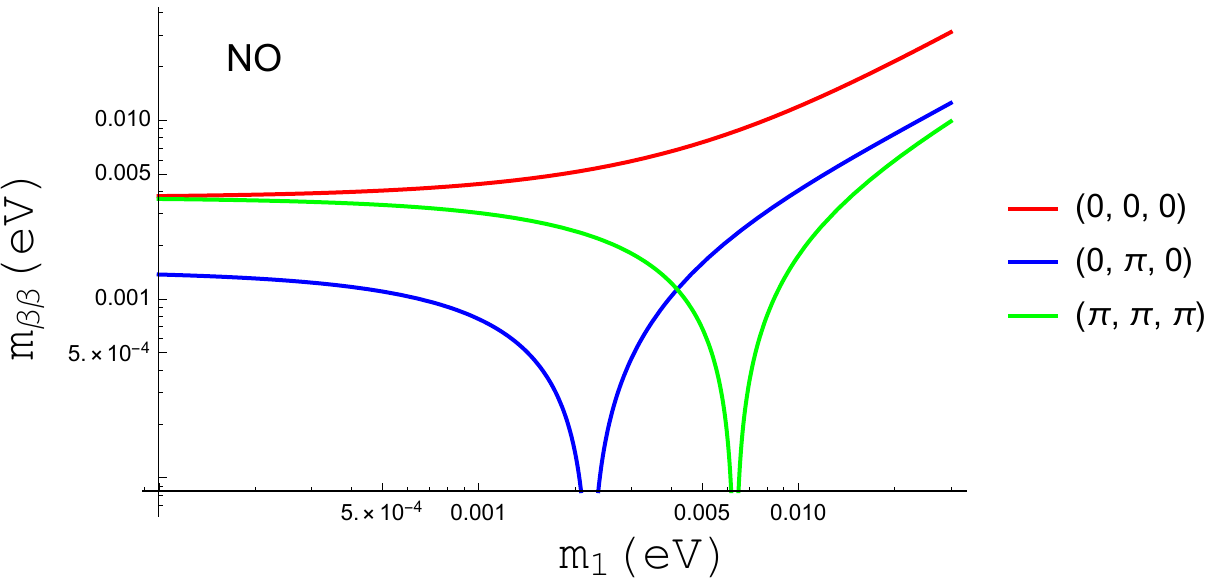}
\end{subfigure}
\hfill 
\begin{subfigure}
\centering
\includegraphics[width=0.4925\textwidth]{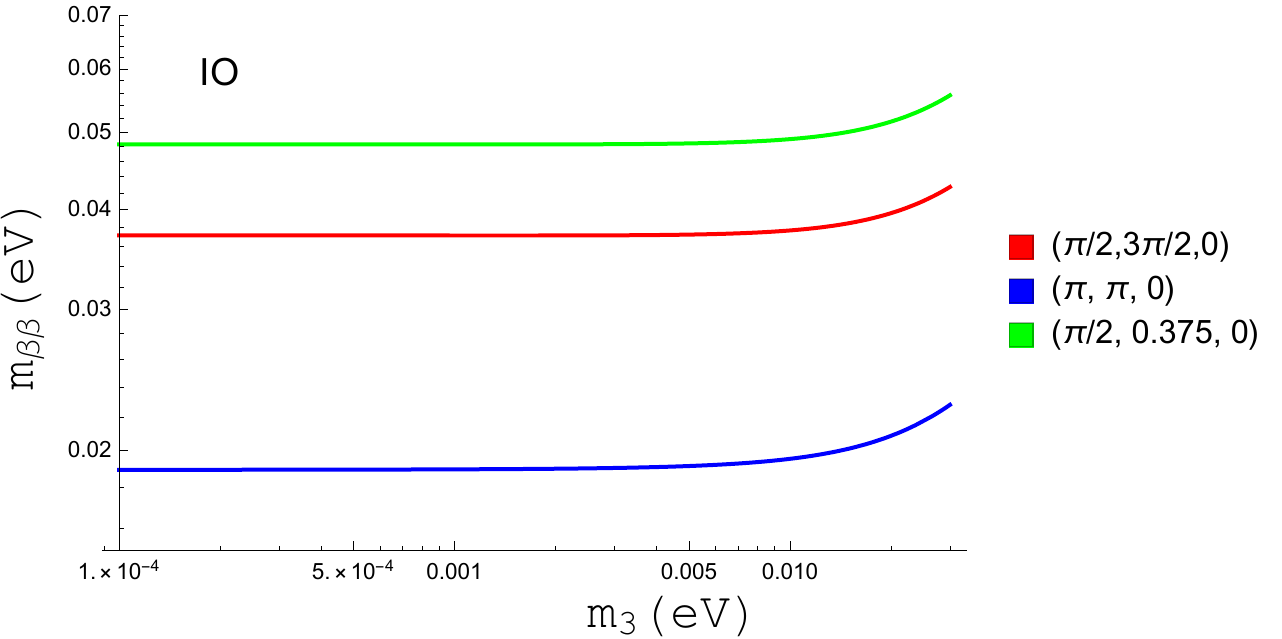}
\end{subfigure}
\caption{Expected values of the $m_{\beta\beta}$ parameter relevant for neutrinoless double beta decay experiments for different sets of neutrino $\mathcal{CP}$ phases ($ \delta_{\mathcal{CP}} $, $ \alpha_{21} $, $ \alpha_{31} $) and varying lightest neutrino mass, for NO (Left) and IO (Right).  All other parameters are given by the best fit parameters in Table \ref{nufit_params}.}
\label{0nubb}
\end{figure}

In Figure \ref{0nubb}, we depict the expected neutrinoless double beta decay signatures embodied by the $m_{\beta\beta}$ mass parameter for different parameter sets of the neutrino $\mathcal{CP}$ phases for varying lightest neutrino mass, in  both the NO and IO schemes. In calculating this result, we have assumed that the contributions from diagrams with  virtual $W^\pm $ and  $ \Delta $ exchange are negligible, which is valid for the range of $ y_{ee} $ and $ m_\Delta $ parameters we consider. The IO scenario offers the best opportunity for measurement of neutrinoless double beta decay. It is clear that the measurement of $ m_{\beta\beta} $ will provide a complementary test to the projected LFV experimental limits depicted above, and thus will be an important step in the confirmation of the existence of the Type II Seesaw mechanism.


\section{Conclusions and Future Prospects}
\label{Conc}

The Type II Seesaw Leptogenesis scenario offers a well-motivated and natural framework in which to simultaneously explain the observed baryon asymmetry of the universe, the origin of the neutrino masses, and the inflationary setting. Importantly, the associated triplet Higgs leads to many phenomenological implications that allow for connecting early universe dynamics to terrestrial experiments. In this work, we have established the unique role that LFV decay processes will play in testing and potentially discovering the components of this scenario. Note that, many of the results in this work are applicable to the Type II Seesaw mechanism in general, even if it is not responsible for Leptogenesis through the scenario we describe. 

These results demonstrate the complementary nature of searches for different lepton flavour violation processes, and their necessity in determining the nature of the neutrino sector. Important dependencies and features are also exhibited for varying $ \Delta $ mass, providing unique simultaneous tests of the triplet Higgs properties. Neutrinoless double beta decay experiments will play an important role, in concert with LFV experiments, to pinpoint the neutrino $\mathcal{CP}$ phases and mass ordering. Thus, it is integral to consider each of these experiments to maximise the possibility for discovery of the neutrino generation mechanism.

The constraints on the maximal neutrino Yukawa coupling and $ m_\Delta $ parameter provided by the LFV experiments are integral to understanding the possible running of the various scalar couplings present in this scenario. These experiments subsequently probe the allowed parameter space that ensures vacuum stability up to the Planck scale, and that lead to inflationary observables consistent with CMB observations. These parameters also determine which of the Higgs' is the dominant component of the inflaton, and subsequently which dimension five lepton violating interaction leads to Leptogenesis. Thus, it is necessary to conduct a combined analysis of the LFV constraints in combination with the requirements for vacuum stability and successful inflation.

In summary, the Type II Seesaw Leptogenesis model exhibits a long list of testable features that allow it to be probed by current and future experiments. This unique combination of phenomenological implications includes the inflationary observables in the Cosmic Microwave Background, Gravitational Waves \cite{Figueroa:2017vfa,Caprini:2018mtu,Hazumi:2019lys}, dominance of leptonic decays of the triplet Higgs at collider searches,  neutrinoless double beta decay, and as discussed in detail in this paper - Lepton Flavour Violating decay processes.


\acknowledgments
NDB was supported by IBS under the project code, IBS-R018-D1. The work of STP was supported in part by the European Union's Horizon 2020 research and innovation programme under the Marie Sk\l{}odowska-Curie grant agreement No.860881-HIDDeN, by the Italian INFN (Trieste section), by the INFN  program on Theoretical Astroparticle Physics and by the World Premier International Research Center Initiative (WPI Initiative, MEXT), Japan. We would like to express special thanks to the Mainz Institute for Theoretical Physics (MITP) of the Cluster of Excellence PRISMA$ ^{+} $ (Project ID 39083149), for its hospitality and support.

\bibliographystyle{JHEP}
\bibliography{bibly}
\end{document}